\documentclass[12pt]{article}
\usepackage[utf8]{inputenc}
\usepackage{graphicx}
\usepackage{amsmath}
\usepackage{hyperref}
\usepackage{cite}
\usepackage{authblk} 
\usepackage{enumitem}

\usepackage{stackengine}
\usepackage[a4paper, margin=1in]{geometry} 
\usepackage{accents}
\usepackage{amssymb} 
\usepackage{ulem}  
\usepackage{bm}
\usepackage{calc,tikz}
\usepackage{titlesec}
\usepackage{float}    
\usepackage{caption}  
\usepackage{booktabs, adjustbox,capt-of,lipsum,afterpage}
\usepackage{multirow}
\usepackage{array, ragged2e, makecell}
\usepackage{chngcntr}
\usepackage{subcaption}
\usepackage{listings}

\titleformat{\section}
{\normalfont\Large\bfseries}{\thesection}{1em}{}

\titleformat{\subsection}
{\normalfont\normalsize\bfseries}{\thesubsection}{1em}{}

\title{\textbf{Some Results on Point Estimation of the Association Parameter of a Bivariate\\ Frank Copula}}

\date{}

\author[1,2,3,6]{\fontsize{11.5}{14}\selectfont Yen-Anh Thi Pham}
\author[2,4,6]{\fontsize{11.5}{14}\selectfont Huynh To Uyen}
\author[1,2,3,5,6,*]{\fontsize{11.5}{14}\selectfont Nabendu Pal}

\affil[1]{\fontsize{11.5}{14}\selectfont Department of Applied Mathematics, Faculty of Applied Science, Ho Chi Minh City University of Technology (HCMUT), VNU-HCM, Ho Chi Minh City, Vietnam}

\affil[2]{\fontsize{11.5}{14}\selectfont Vietnam National University Ho Chi Minh City, Linh Trung Ward, Thu Duc City, Ho Chi Minh City, Vietnam}

\affil[3]{\fontsize{11.5}{14}\selectfont Faculty of Mathematics and Statistics, Ton Duc Thang University, Ho Chi Minh City, Vietnam}

\affil[4]{\fontsize{11.5}{14}\selectfont Department of Mathematics and Economic Statistics, University of Economics and Law, Ho Chi Minh City, Vietnam}

\affil[5]{\fontsize{11.5}{14}\selectfont Department of Mathematics, University of Louisiana at Lafayette, Lafayette, Louisiana, USA}

\affil[6]{\fontsize{11.5}{14}\selectfont Vietnam Institute for Advanced Studies in Mathematics (VIASM), Hanoi, Vietnam}

\begin{document}
	\label{firstpage}
	\maketitle

\renewcommand{\thefootnote}{} 
\footnotetext{
	\begin{flushleft} 
		{\bf Keywords:} Maximum likelihood estimation, method of moment estimation, bias, mean squared error, asymptotic distribution. 
		
		\vspace{0.5\baselineskip} 
		\textbf{MSC 2020 Subject Classifications:} 62F10, 62E20, 62H05 
		
		\vspace{0.5\baselineskip} 
		$^*$ {\bf Corresponding author:} nabendupal@gmail.com; nabendu.pal@tdtu.edu.vn; nabendu.pal@louisiana.edu
	\end{flushleft}
}
\renewcommand{\thefootnote}{\arabic{footnote}} 
\vspace{-1cm}  
	\begin{abstract}
		This work deals with estimation of the association parameter of a bivariate Frank Copula in a comprehensive way. Even though Frank Copula is a member of Archimedean class of copulas, and has been widely used in finance, relatively little attention has been paid to its association parameter from a statistical inferential point of view. Most of the existing works which have used Frank Copula have focused on estimating the parameter computationally, and then proceeded with its application in the applied fields, mostly in finance. Here, in this investigation, we have looked at the point estimation of the association parameter in a comprehensive manner, and studied three estimators in terms of bias, mean squared error (MSE), relative bias and relative MSE. It has been noted that in the neighborhood of zero, the method of moment estimators (MMEs) do perform well compared to the maximum likelihood estimator (MLE), even though the latter has the best overall performance. Further, in terms of bias, MMEs and MLE have opposite behavior. However, some of  our results do not match with those reported by Genest (1987) \cite{Genest1987}. Nevertheless, this study complements Genest's (1987)\cite{Genest1987} expository work, and provides some interesting insights into the behaviors of three point estimators including the MLE whose asymptotic behavior holds pretty well, as we have found, for $n\ge 75$.
	\end{abstract}
	
	

	
\section{Introduction}
	\subsection{Preliminaries}
	In the univariate case we have several distributions to choose from when the data shows a skewed pattern even though the normal (or Gaussian) distribution is widely used, often blindly. However, in a multidimensional setup, either by omission or by commission, the multivariate normal distribution (MND) appears to be the default choice in modelling the data under study. There are definitely certain advantages in using the MND, such as: (i) The sampling distributions of the sample mean and the sample dispersion matrix are independent and well-known; (ii) Testing statistical hypotheses on the population mean as well as the dispersion matrix are well-studied and well-understood. However, many of the standard multivariate statistical inferential results fall apart if the data do not follow the MND. Therefore, it is very important that the applied researchers must test the validity of MND for a given dataset before proceeding with subsequent inferences.

	There are plenty of instances, ranging from ecological studies to engineering, from gene expression to financial studies, where we see deviations from the normality assumption (and for such a dataset see Chatterjee, R. (2022)) \cite{Chatterjee2022}. This can be easily verified often (but not always) by looking at the marginal relative frequency histograms of the components of the multivariate data under study. If the multivariate data under study is assumed to be MND, then the marginals must be univariate normal (but not the other way around). Yet, there are plenty of examples, where the multivariate data do not follow the MND, and hence one must look for alternative ways to model such data. One option is to develop a suitable non-normal multivariate distribution starting with the observed marginals, and then combine them by using a suitable copula. Based on the famous theorem put forward by Sklar (1959) \cite{Sklar1959}, the copula approach presents a much-needed alternative to the usual ``top-down" approach of the MND, which can work as a ``bottom-up" approach by combining the non-normal marginals in a suitable way.\\
	
	In the following subsection, we provide a brief description of the copula theory in general, and the Frank Copula in particular, which is the focus of this work.
		
\subsection{A Brief Introduction to the Copula Theory}
Consider a $p$-dimensional random vector $\bm{X} = (X_1, X_2,..., X_p)^{'}$ with absolutely continuous marginals where the marginal cdf of $X_i$ is the given by $F_i(x_i) = P(X_i \leq x_i)$. By using the probability integral transform to each component, i.e., $U_i = F_i(X_i)$, the random vector $\bm{U} = (U_1, U_2,...,U_p)^{'}$ has Uniform $(0,1)$ marginals. The Copula of $\bm{X}$ is defined as the joint cdf of $\bm{U} = (U_1, U_2,...,U_p)^{'}$ given as
		\begin{align*}
		C(\bm{u}) & = C(u_1, u_2, \ldots, u_p) = P(U_1 \leq u_1, \ldots, U_p \leq u_p) \\
			& = P(X_1 \leq F_1^{-1}(u_1), \ldots, X_p \leq F_p^{-1}(u_p)) \tag{1.1}
		\end{align*}
Sklar's (1959) \cite{Sklar1959} pathbreaking result provides the theoretical basis for the application of copulas which states that the joint \textit{cdf} \(F(\bm{x})\) of $ \bm{X} $  can be represented in terms of its marginals, i.e., \(F_i(x_i)\) for \(1 \leq i \leq p\), and a suitable copula \(C\), i.e.,
	\[
		F(\bm{x}) = C(F_1(x_1), F_2(x_2), \dots, F_p(x_p)).  \tag{1.2}
	\]
Also, since we assume that $\bm{X}$ is absolutely continuous, the joint \textit{pdf} $ f(\bm{x}) $ of  $ \bm{X} $ can be expressed as 
	\[
		f(\bm{x}) = f(x_1, x_2, \ldots, x_p) = c(F_1(x_1), F_2(x_2), \ldots, F_p(x_p)) f_1(x_1) f_2(x_2) \ldots f_p(x_p) \tag{1.3}
	\]
Where \( f_i(x_i) \) is the \textit{pdf} of \( X_i \) at \( x_i \), and $c(\bm{u}) = C^{(\bm{u})} (\bm{u})$, where 
	\[
		c(\bm{u}) = C^{(\bm{u})} (\bm{u}) = \frac{\partial}{\partial u_p} \frac{\partial}{\partial u_{p-1}} \ldots \frac{\partial}{\partial u_1} C(u_1, u_2, \ldots, u_p). \tag{1.4}
	\]
	
There are many well-known families of copulas discussed in the literature, and for a nice overview one can see Nelsen (2007) \cite{Nelsen2007}. A large class of copulas (which is essentially a collection of subclasses of copulas) is called Archimedean if it admits the representation 
	\[
		C(\bm{u}) = C(u_1, u_2, \ldots, u_p \mid \theta) = \psi_{\theta}^{-1}(\psi_{\theta}(u_1) + \psi_{\theta}(u_2) + \ldots + \psi_{\theta}(u_p)) \tag{1.5}
	\]
Where $\psi_{\theta} : [0,1] \rightarrow [0, \infty) $ is a continuous, strictly decreasing and convex function such that $\psi_\theta(1) = 0$. The parameter $\theta$, often called the association parameter, takes values over the parameter space $\Theta$, and $\psi$ is called the generator function. \\
	
Some well-known copulas of the above Archimedean family (1.5), in the bivariate case $(p=2)$, are: 
	\begin{enumerate}[label=(\roman*), align=left]
		
		\item Clayton Copula: generator $\psi_\theta(u)=(u^{-\theta}-1)/\theta$, $\theta \in \Theta$ = $[-1, \infty) \setminus \{0\}, \text{ and } C(\bm{u}) = \left[ \mathit{max} \left\{ u_1^{-\theta} + u_2^{-\theta} - 1, 0 \right\} \right]^{-1/\theta} $
		
		\item Gumbel Copula: generator $\psi_\theta(u)= (-\mathit{ln} u)^\theta$, $\theta \in \Theta = [1,\infty)$, \text{ and } $C(\bm{u}) = \mathit{exp} {[ -\left\{(-\mathit{ln} u_1)^\theta + (-\mathit{ln} u_2)^\theta \right\}^{1/\theta}]}.$
				
		\item Joe Copula: generator $\psi_\theta(u)= -\mathit{ln}(1-(1-u))^\theta$, $\theta \in \Theta $ = $[1,\infty), \text{ and } C(\bm{u}) $ = $ 1-\left\{(1-u_1)^\theta + (1-u_2)^\theta - (1-u_1)^\theta(1-u_2)^\theta\right\}^{1/\theta}.$
	\end{enumerate}
	
	Some well-known copulas which are not members of the Archimedean class are Gaussian copula, Farlie-Gumbel-Morgenstern copulas, etc.
		
\subsection{Frank Copula and Some Basic Properties: The Bivariate Case}
	
	An important member of the aforementioned Archimedean class of copulas is the Frank copula which has the generator
	\[
		\psi_\theta(u) = - \mathit{ln} \{(\mathit{exp}(-\theta u)-1) / (\mathit{exp} (-\theta)-1)\},  \tag{1.6}
	\]
with the parameter space $\Theta = \mathbb{R} \setminus \{0\}$ with the corresponding bivariate copula
	\[
		C(u_1,u_2|\theta) = -(1/\theta) \mathit{ln} \left[ 1+ \left\{\mathit{exp}(-\theta u_1) - 1\right\} \left\{\mathit{exp}(-\theta u_2) - 1\right\} / \left\{\mathit{exp}(-\theta) - 1\right\} \right]. \tag{1.7}
	\]
	
In typical bivariate applications when we have bivariate data on $\bm{X} = (X_1, X_2)^{'}$, and it is assumed that the distribution of $\bm{X}$ can be expressed in terms of the Frank copula, the actual \textit{pdf} of $\bm{X} $ will take the form $(1.3)$ where $c(u_1, u_2|\theta) = \partial^2 C(u_1, u_2)/ \partial u_1 \partial u_2, $ and $f_i(x_i)$ being the marginal \textit{pdf} of $X_i, i=1,2$, which may have their own parameters (associated with the corresponding marginal distributions) apart from the association parameter $\theta$ of the copula. But to keep this investigation relatively simple we either assume that $f_i(x_i)$ are either completely known, or $F_i(x_i)$ can be replaced by $\hat{F}_i(x_i)$, where $\hat{F}_i(.)$ is the empirical \textit{cdf} of $X_i$, obtained from the observations on the $ i \underbar{\textit{th}} $ component $X_i$, i.e., $X_{i1}, X_{i2},...,X_{in}, i=1,2$. The usual form of $\hat{F}_i(t)$ is given as 
	\[
		\hat{F}_i(t) = (1/n) \textstyle \sum_{j=1}^n I(X_{ij} \leq t),  \tag{1.8}
	\]
where $I(.)$ indicates the indicator function. Sometimes a little adjustment is done in (1.8) to take care of the extreme values 0 and 1 by using
	\[
 		\hat{F}_{i(adj)}(t) = \{\textstyle \sum_{j=1}^n I(X_{ij} \leq t) + 0.5 \}/(n+1). \tag{1.9}
	 \]
	The idea of using the empirical \textit{cdf} (or its adjusted version) is that one can use $U_i = F_i(X_i) \simeq \hat{F}_i(X_i)$ (or $\hat{F}_{i(adj)}(X_i)$) which, for moderately large $n$, would be approximately (but not exactly) Uniform $(0,1)$. Therefore, we can pretend that we have data on $\bm{U} = (U_1, U_2)$ (in the bivariate case) with joint \textit{pdf}
		
	\[
		c(u_1, u_2 \mid \theta) = \frac{\theta (1 - \mathit{exp}(-\theta)) (\mathit{exp}(-\theta(u_1 + u_2)))} {\left\{ \mathit{exp}(-\theta u_1) + \mathit{exp}(-\theta u_2) - \mathit{exp}(-\theta)-\mathit{exp}(-\theta(u_1 + u_2))\right\}^2}, \tag{1.10}
	\]
where $\bm{u} = (u_1, u_2) \in (0,1)\otimes(0,1)$ and $\theta \in \Theta = \mathbb{R} \setminus \{0\}. $ Note that while (1.10) is the joint \textit{pdf}, the expression of $C(u_1, u_2|\theta)$ in (1.7) is the joint \textit{cdf} of the uniform marginals of $U_1$ and $U_2$. From now on we call (1.10) (or, equivalently (1.7)) as the \textbf{Standard Bivariate Frank Copula Distribution with parameter $\theta$ $\mathbf{(SBFCD(\theta))}$.}\\
	
The special situation of $\theta = 0$ in (1.10) requires a deft handling as a direct plug-in in (1.10) or in (1.7) yields $0/0$. So, to understand $SBFCD(\theta)$ at $\theta = 0$ one should take limit $\theta \to 0$ either by L'Hospital's rule or by using the approximations (as $\theta \to 0$): $\mathit{exp}(u)-1 \approx u$ and $\mathit{ln}(1+u) \approx u$ as $u \to 0$ and this yields $c(u_1,u_2|\theta \to 0) = 1 $  or $C(u_1,u_2|\theta \to 0) = u_1 u_2$, i.e., the components are independent.\\
	
Our comprehensive study on the association parameter $\theta$ requires a good amount of simulations, and hence it is important to execute the data generation and computations within a reasonable amount of time. However, the existing R code commands [Given in the Appendix A.1] appears to be time consuming thereby slowing down the simulation. Hence, in the following, we provide the data generation technique through four simple steps which appears to work much faster. \\

As mentioned earlier, the aforementioned Frank Copula has been applied extensively in finance to study the interdependencies among various assets. For a detailed discussion on this topic see Jouanin et al. (2004) \cite{Jouanin2004} or Fairuz (2024) \cite{Fairuz2024} and further references therein. However, none of these papers, especially the latter one, does not deal with the in-depth study we are providing here about the association parameter estimation.
\subsection{ Data Generation from SBFCD($\theta$)}

The following data generation algorithm has been used for our simulation study, and it is easy to implement in R package.
	
	Fix $\theta \in \Theta =  \mathbb{R} \setminus \{0\}.$ (For $\theta = 0, \bm{U} = (U_1, U_2)$ has independent components $U_i \sim $ Uniform $(0,1), i=1,2.$)\\
	\textbf{Step - 1}: Generate $U_1 = u_1$ from Uniform $(0,1).$\\
	\textbf{Step - 2}: Generate $V = v $ from Uniform $(0,1)$ (independent of $U_1$).\\
	\textbf{Step - 3}: Solve for $u_2$, where
	\[
		u_2 = (-1/\theta) \mathit{ln} \left[ (T_1/T_2) - (A/B)\right],      \tag{1.11}
	\]
where $ A = \mathit{exp}(-\theta u_1) - \mathit{exp}(-\theta),
		B = 1 - \mathit{exp}(-\theta u_1), D = 1 - \mathit{exp}(-\theta), T_1 = D((A/B) + 1) \text{ and } T_2 = v(\mathit{exp}(\theta u_1) - 1)(A + B) + D
		.$\\
	(The above $u_2$ is an observation generated from the conditional distribution of $(U_2|U_1 =u_1)$, where $u_1$ is given in Step - 1.)\\
	\textbf{Step - 4}: The pair $(u_1, u_2)$ thus obtained is an observation of $(U_1, U_2) \sim SBFCD(\theta).$ (Details of the justification of the above Step - 3 is provied in the Appendix A.2.)\\
	
	In case of generating a sample of size $n$, just repeat the above four steps $n$ times to obtain $\bm{U}_j$ = $(U_{1j},U_{2j})^{'}, 1 \leq j \leq n$.\\
	
	To see how $SBFCD(\theta)$ varies for different values of $\theta$, we have generated the following scatterplots for $\theta = -10, -0.1, +0.1, \text{ and } + 10.$ as shown in Figure $1.1.$ 
	
	\begin{figure}[H]
	\centering
	\begin{subfigure}[b]{0.48\textwidth}
		\includegraphics[width=\textwidth, height=6.0cm]{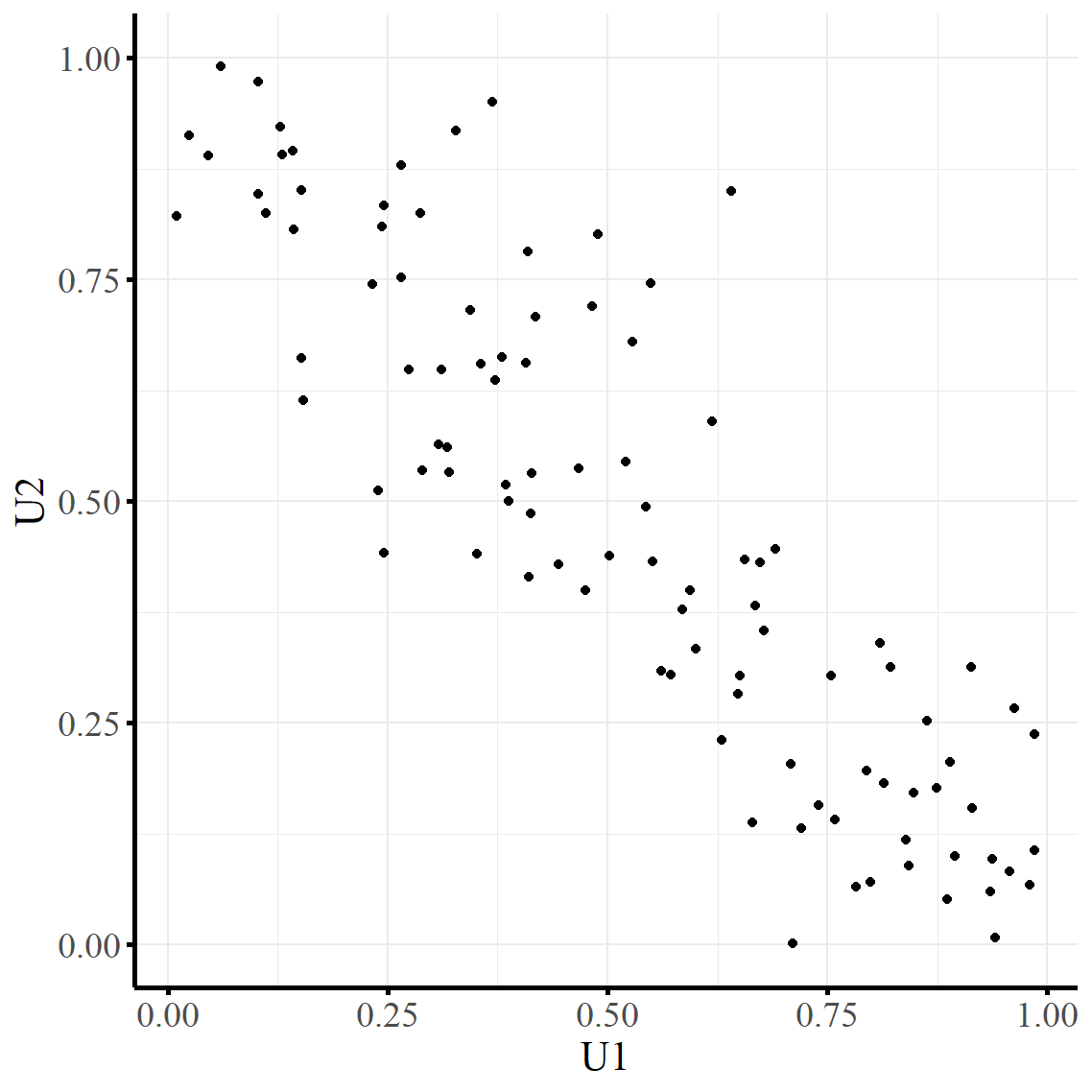}
		\caption{$(\theta = -10)$}
		\label{fig:theta_minus10}
	\end{subfigure}
	\hspace{0.02\linewidth}  
	\begin{subfigure}[b]{0.48\textwidth}
		\includegraphics[width=\textwidth, height=6.0cm]{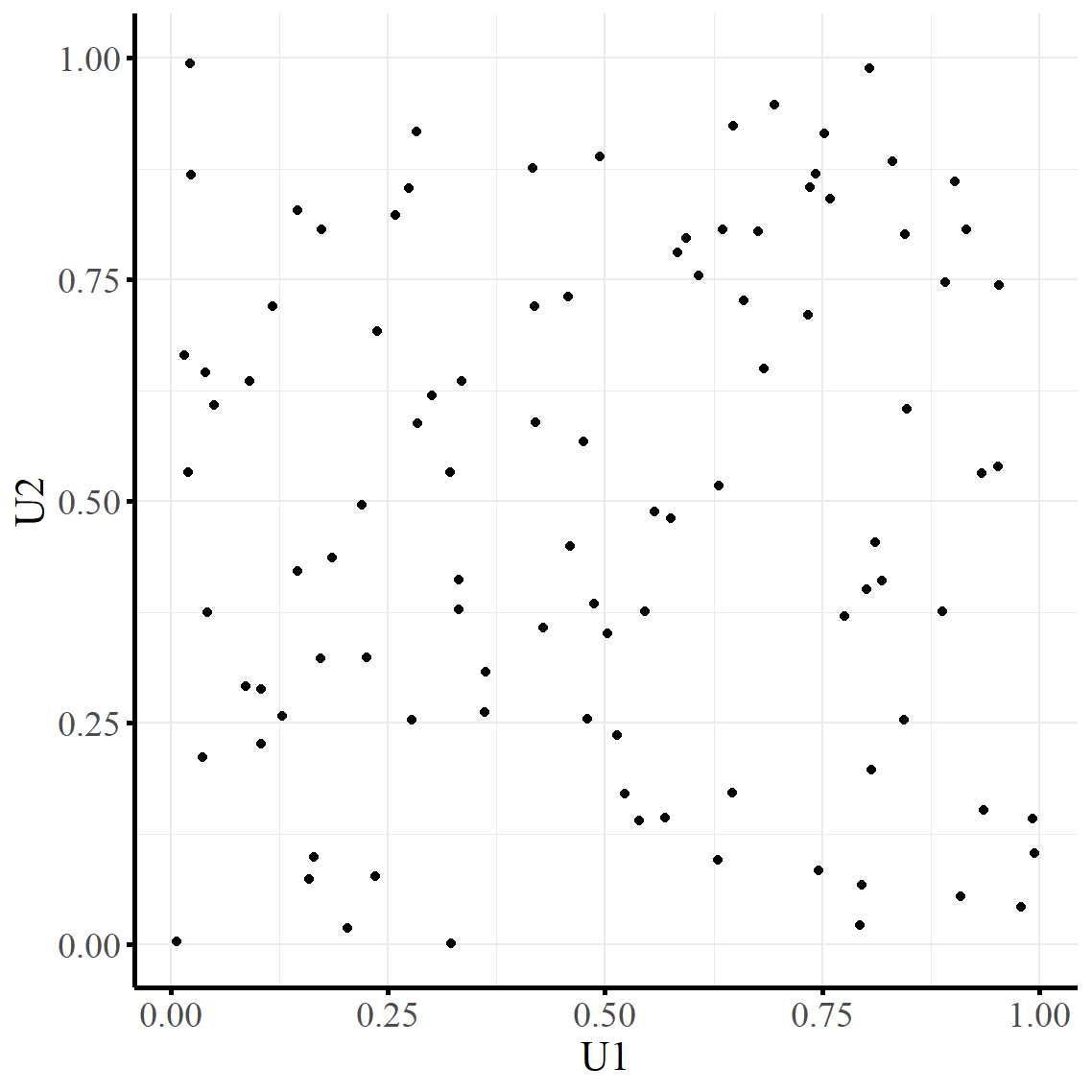}
		\caption{$(\theta = -0.1)$}
		\label{fig:theta_minus0.1}
	\end{subfigure}
	
	\vskip\baselineskip
	
	\begin{subfigure}[b]{0.48\textwidth}
		\includegraphics[width=\textwidth, height=6.0cm]{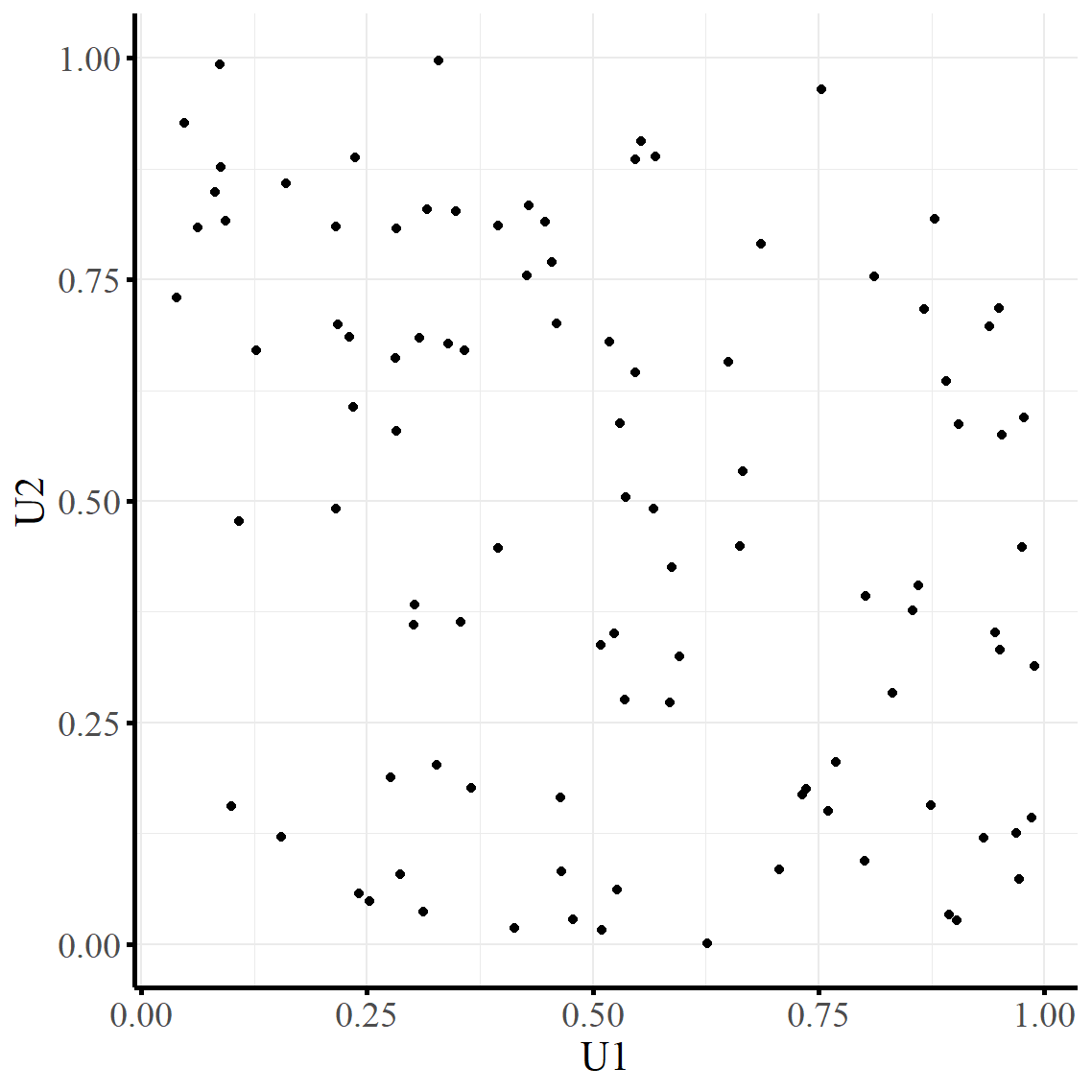}
		\caption{$(\theta = 0.1)$}
		\label{fig:theta_0.1}
	\end{subfigure}
	\hspace{0.02\linewidth}  
	\begin{subfigure}[b]{0.48\textwidth}
		\includegraphics[width=\textwidth, height=6.0cm]{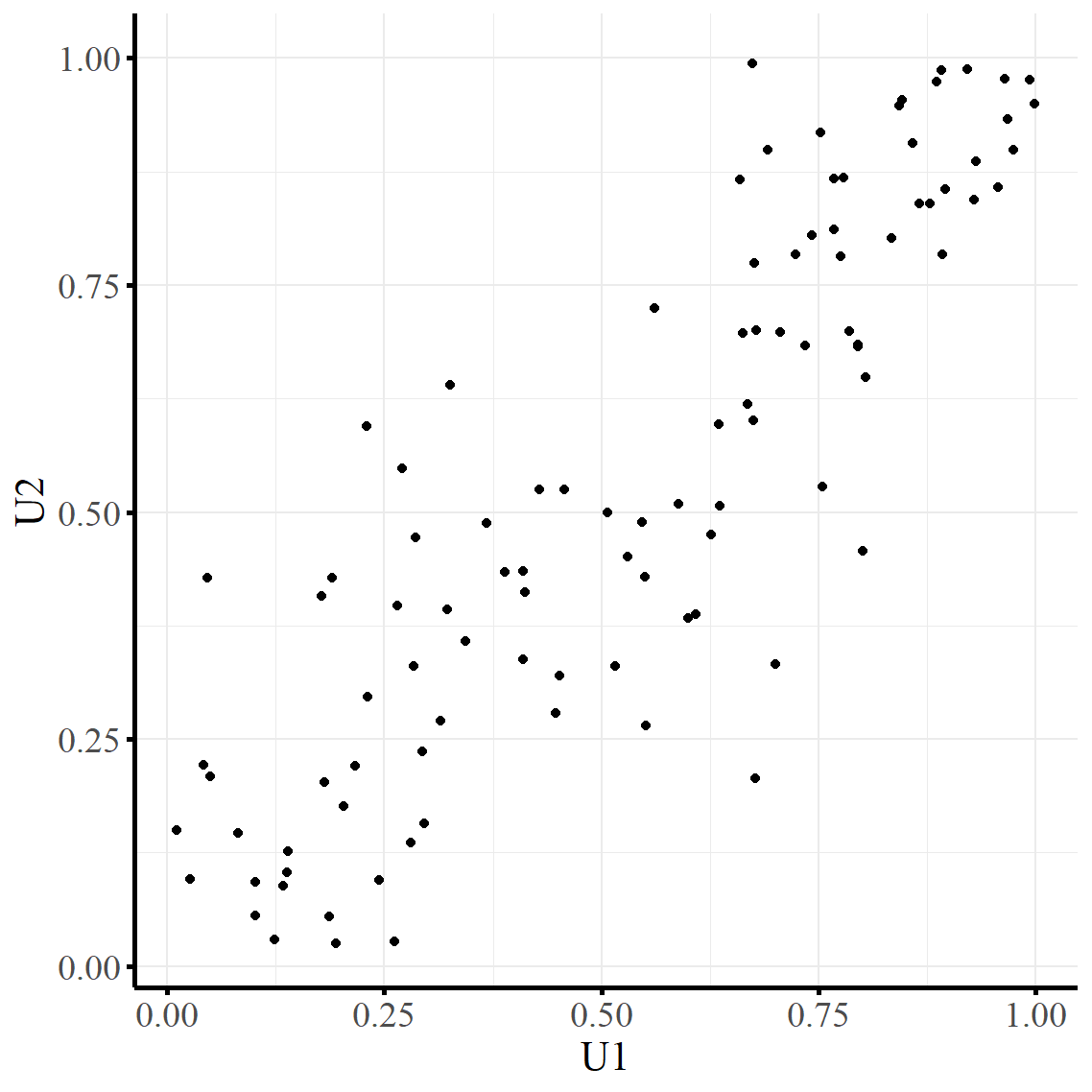}
		\caption{$(\theta = 10)$}
		\label{fig:theta_10}
	\end{subfigure}
	
	\caption{Figure 1.1. Scatter plots of \textit{n} = 100 observations from SBFCD($\theta$)}
	\label{fig:scatterplots}
\end{figure}
As expected, when $\theta$ is extreme (i.e., far from $0$), the discernible patterns in the scatter plots are easily evident. But when $\theta$ is close to $0$, the patterns are nearly indistinguishable. Also, from the above Figure 1.1 it may appear that the mean regression of $U_2$ on $U_1$, i.e., $E(U_2 | U_1)$ is linear in $U_1$, but in essence it is not, rather it can be a curve-linear structure as shown in
Dette et al. (2014) \cite{Dette2014} (see their Figure 1).

\subsection{A Brief Summary of This Work}
\begin{enumerate}[label=(\alph*)] 
	\item In section 2 we have dealt with the existence of maximum likelihood estimator (MLE) of the parameter $\theta$ which is obtained by solving the normal equation $H(\theta) \text{(say)} = 0$ (given in (2.4)). In the process, we also study the behavior of $H(\theta)$ as a function of $\theta \in \mathbb{R}$.
	\item In section 3 we provide the expressions of two method of moment estimators (MMEs) details of which can also be found in Genest (1987) \cite{Genest1987}.
	\item One of the major contributions of this work is Section 4 where we provide  the results of our comprehensive simulation study comparing the three estimators of $\theta$, the MLE and two MMEs. Even though Genest(1987) \cite{Genest1987} provided results of his limited computational study, it has been observed that some of his results are at variance with ours. Among other things it has been observed that even though the MLE performs better in general, near $\theta = 0$ the MMEs can outperform the MLE in terms of MSE (or RMSE). Also, in terms of bias, MLE and MMEs perform in
	opposite directions.
	\item Section 5 is devoted to study the asymptotic behavior of the MLE, and among other things it has been shown that the MLE attains its asymptotic variance fairly well for a minimum sample size of $75$.
\end{enumerate}
\section{Maximum Likelihood Estimation (MLE) of $\theta$}

\subsection{Derivation of the MLE}
	
If a single observation $\bm{U}$ = $(U_1, U_2)^{'} \sim c(u_1, u_2|\theta)$ as in (1.10), the log-likelihood of the single observation is	
	\begin{align*}
		\mathit{ln} c(U_1, U_2 \mid \theta) &= \mathit{ln} \theta + \mathit{ln} (1 - \mathit{exp}(-\theta)) - \theta (U_1 + U_2) \\
		 & - 2 \mathit{ln} \left\{ \mathit{exp}(-\theta U_1) + \mathit{exp}(-\theta U_2) - \mathit{exp}(-\theta) - \mathit{exp}(-\theta (U_1 + U_2)) \right\}. \tag{2.1}
	\end{align*}
	
Define $A_1(U)$ and $A_2(U)$ as follows
		\begin{align*}
			A_1(\bm{U}) &=  \textstyle \sum_{i=1}^2 { U_i \mathit{exp}(-\theta U_i)} - (\sum_{i=1}^2 U_i) \mathit{exp} (-\theta \sum_{i=1}^2 U_i) - \mathit{exp}(-\theta); \\
			A_2(\bm{U}) &=  \textstyle \sum_{i=1}^2 \left\{ \mathit{exp}(-\theta U_i) \right\} - \mathit{exp} (-\theta \sum_{i=1}^2 U_i ) - \mathit{exp}(-\theta).
		\end{align*}
Then
	\begin{equation}
		\frac{\partial}{\partial \theta} \mathit{ln} c(U_1, U_2 \mid \theta) = (1/\theta)+ \left( \mathit{exp}(\theta) - 1\right)^{-1} - \left(\textstyle \sum_{i=1}^2 U_i\right) + 2 \left(A_1/A_2 \right). \tag{2.2}
	\end{equation}
	
The above expression will be useful in deriving the MLE as well as the `Fisher Information Per Observations'(FIPO). \\
	
	Based on a sample of $n$ observations $\bm{U}_j$ = $(U_{1j}, U_{2j})^{'}, 1 \leq j \leq n, $ the log-likelihood function is 
	
	\begin{equation}	
		l(\theta) = \textstyle \sum_{j=1}^n \mathit{ln} c(U_{1j}, U_{2j} \mid \theta),   \tag{2.3}
	\end{equation}
and the MLE, denoted by $\hat{\theta}_{ML}$, is found by setting $l^\prime (\theta) = 0$, i.e., by solving the normal equation
	\[
		H(\theta) (\text{say}) = \left( 1/\theta\right) + \left( \mathit{exp}(\theta) - 1 \right)^{-1} - \left( \textstyle \sum_{i=1}^2 \overline{U}_i \right) + (2/n) \left\{ \textstyle \sum_{j=1}^n A_1(\mathbf{U}_j)/A_2(\mathbf{U}_j) \right\} = 0. \tag{2.4}
	\]
Where $ \overline{U}_i =  (\sum_{j=1}^n U_{ij}/n), i=1,2,$ \text {and } $A_i(\bm{U}_j)$ is the expression of $A_i(\mathbf{U})$ given after (2.1) with $\bm{U}$  being replaced by $ \bm{U}_j, 1 \leq j \leq n, i=1,2.$
	
	As a demonstration, we have generated a random sample of size $n=25$ observations from $ \mathit{SBFCD(\theta)}$ with $\theta = 1$ [given in Appendix A.3 (Table A.1)]. The following Figure 2.1 shows the plots of $l(\theta)$ and $H(\theta)$ against $\theta$ where a small neighborhood of $(-0.01, +0.01)$ around $0$ has been excluded. 
	\begin{figure}[H]
		\centering
		\captionsetup[subfigure]{labelformat=empty}
		
		\begin{subfigure}[b]{0.45\linewidth}
			\centering
			\includegraphics[width=\linewidth]{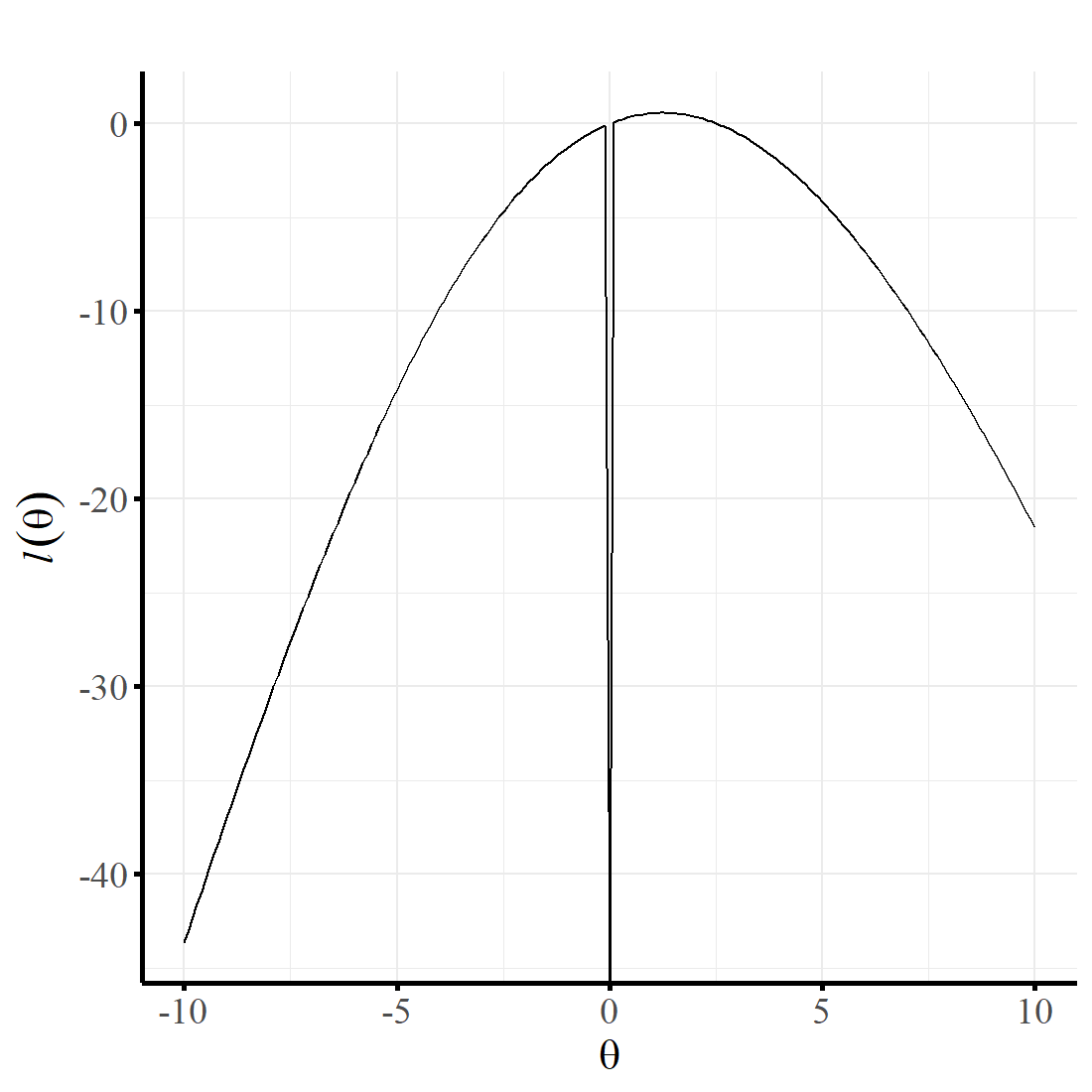}  
			\caption{Plot of $l(\theta)$ against $\theta$}
			\label{fig:plot_l_theta}
		\end{subfigure}
		\hspace{0.02\linewidth}  
		\begin{subfigure}[b]{0.45\linewidth}
			\centering
			\includegraphics[width=\linewidth]{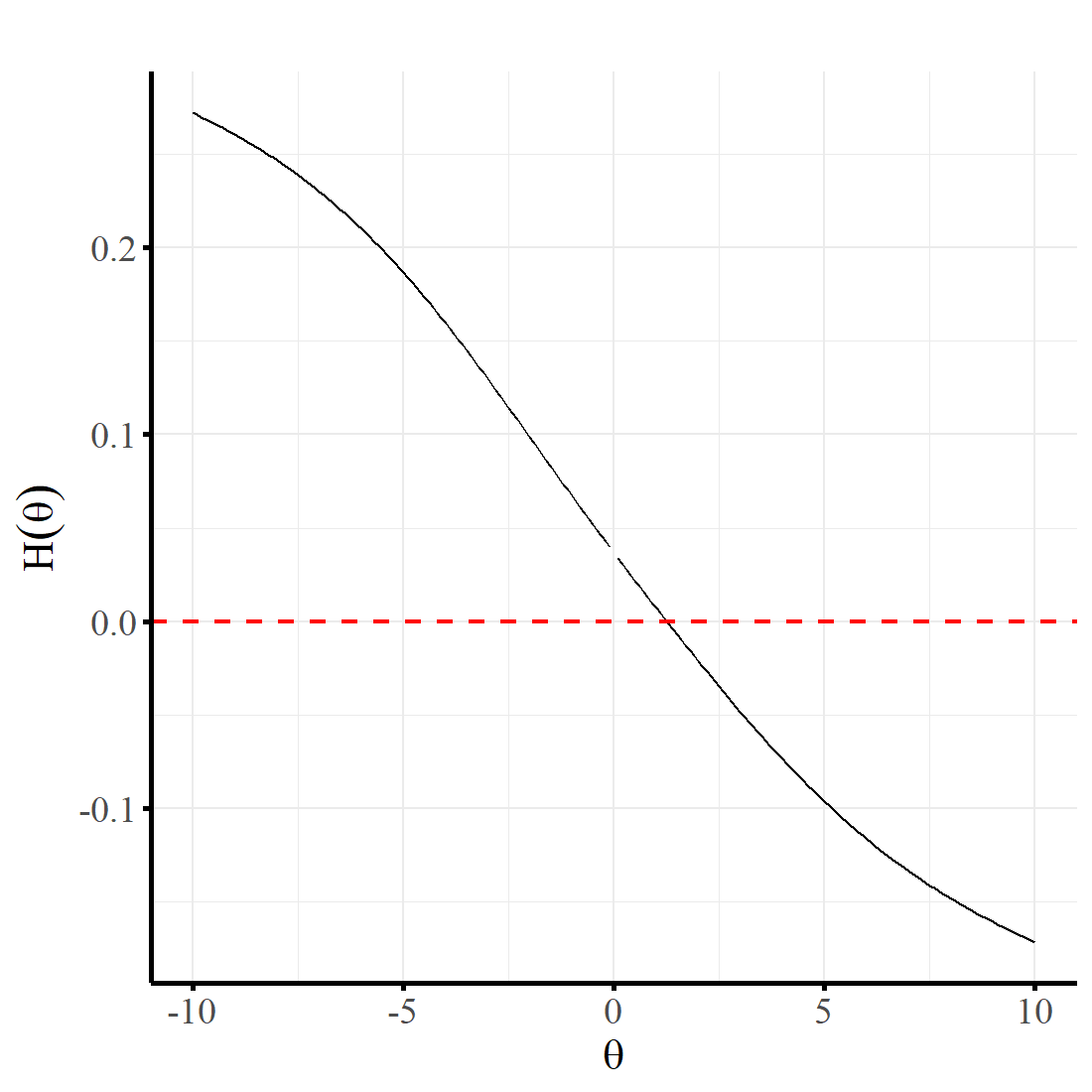}  
			\caption{Plot of $H(\theta)$ against $\theta$}
			\label{fig:plot_H_theta}
		\end{subfigure}
		
		\caption{Figure 2.1. Plots of $l(\theta)$ and $H(\theta)$ against $\theta$, $\theta \in \mathbb{R} \setminus \{0\}$.}
		\label{fig:plots_l_H_theta}
	\end{figure}
	Even though the solution of (2.4) ought to be found by numerical means to obtain $\hat{\theta}_{ML}$, it is important to study the behavior of $H(\theta)$ for a given dataset to ensure that the MLE exists and this is shown in the next subsection.
	
\subsection{Behavior of the function $H(\theta)$ in (2.4) and Existence of the MLE}

	\textbf{(A) Behavior of} $\mathbf{H(\theta)}$ \textbf{as} $\mathbf{\theta \to \infty}$: For the sake of convenience, define $H_j(\theta)$ as
	\[
	H_j(\theta) = A_1(\bm{U}_j)/A_2(\bm{U}_j),
	\]
which is the expression inside the sum in the last term of $H(\theta)$ in (2.4). Then $H_j(\theta)$ can be expressed as 
	\[
		H_j(\theta) = \frac{U_{1j} \mathit{exp}(\theta U_{2j}) + U_{2j} \mathit{exp}(\theta U_{1j}) - \mathit{exp}\left(\theta(U_{1j} + U_{2j} - 1)\right) - \sum_{i=1}^{2} {U_{ij}}}{\mathit{exp}(\theta U_{2j}) + \mathit{exp}(\theta U_{1j}) - \mathit{exp}\left(\theta(U_{1j} + U_{2j} - 1)\right) - 1}.
	\]
W.l.g. further assume that $U_{1j} > U_{2j}$. Then, after multiplying both the numerator, and the denominator by $exp(-\theta U_{ij})$, we get
	\[
		H_j(\theta) = \frac{U_{1j} \mathit{exp}\left(\theta \left(U_{2j} - U_{1j}\right)\right) + U_{2j} - \mathit{exp}\left(\theta \left(U_{2j} - 1\right)\right) - \left(\sum_{i=1}^{2} U_{ij}\right) \mathit{exp}\left(-\theta U_{1j}\right)}{\mathit{exp}\left(\theta \left(U_{2j} - U_{1j}\right)\right) + 1 - \mathit{exp}\left(\theta \left(U_{2j} - 1\right)\right) - \mathit{exp}\left(-\theta U_{1j}\right)}
	\]
As $\theta \to \infty$, $ \mathit{exp} \left( \theta \left( U_{2j} - U_{1j}\right)\right) \to 0$ as $(U_{2j} - U_{1j}) < 0 $, $ \mathit{exp} \left( \theta \left( U_{2j} - 1 \right)\right) \to 0 $ since $U_{2j} < 1 $, and $\mathit{exp} \left( -\theta U_{1j}\right) \to 0 $ since $U_{ij}>0$. Therefore, $H_j(\theta) \to U_{2j}$. A similar argument can be presented to show that $H_j(\theta) \to U_{1j}$ as $\theta \to \infty$ if $U_{1j} < U{2j}$. Thus, irrespective of the order of $U_{1j}$ and $U_{2j}, H_j(\theta) \to  \mathit{min}(U_{1j}, U_{2j}).$ So, as $\theta \to \infty, (1/\theta) \to 0, (\mathit{exp}(\theta) -1)^{-1} \to 0$, and $ (2/n) \{ \sum_{j=1}^n A_1(\bm{U}_j)/A_2(\bm{U}_j)\} \to (2/n)\{\sum_{j=1}^n \mathit{min}(U_{1j}, U_{2j})\}$. 
Hence, $H(\theta) \rightarrow -(\overline{U}_1 + \overline{U}_2) + (2/n) \sum_{j=1}^n \mathit{min}(U_{1j}, U_{2j})$ which is \textbf{strictly negative}. [Note that $-(\overline{U}_1 + \overline{U}_2) = -( \sum_{j=1}^n (U_{1j}+ U_{2j}))/n = -(\sum_{j=1}^n \mathit{min}(U_{1j}, U_{2j}) + \sum_{j=1}^n \mathit{max} (U_{1j}, U_{2j}))/n$. Therefore, $H(\theta) \rightarrow - (\sum_{j=1}^n \mathit{max}(U_{1j}, U_{2j}))/n + (\sum_{j=1}^n \mathit{min}(U_{1j}, U_{2j}))/n < 0$.]\\
	
\noindent \textbf{(B) Behavior of} $\mathbf{H(\theta)}$ \textbf{as} $\mathbf{\theta \rightarrow 0^+}$ \textbf{(as well as} $\mathbf{ \rightarrow 0^-)}$: Using the argument that near $x = 0$, $\mathit{exp}(x) - 1 \approx x \approx \mathit{ln}(x + 1)$, it can be shown that $H(\theta) \approx 2 - (\overline{U}_1 + \overline{U}_2) + 4(\overline{U_1U_2}) > 0$, where $\overline{U_1U_2} = (\sum_{j=1}^n U_{1j} U_{2j}/n)$.\\

The fact that near $\theta = 0^+ $, the expression $H(\theta) = 2 - (\overline{U}_1 + \overline{U}_2) + 4(\overline{U_1U_2}) > 0$ shows that this behavior holds as well when $\theta \to 0^-$.\\
	
\noindent \textbf{(C) Behavivor of} $\mathbf{H(\theta)}$ \textbf{as} $\mathbf{\theta \to -\infty} \textbf{:}$.
To see this case, for the sake of convenience, we write $\theta = -\delta$. So, essentially we need to see what happens to $H(\theta) = H(-\delta)$ as $\delta \to \infty$. We can express $H(-\delta)$ as follows
	\begin{align*}
		H(-\delta) &= -\left(1/\delta\right) - \left(\mathit{exp}(-\delta) - 1\right)^{-1} - \left(\overline{U}_1 + \overline{U}_2\right) + (2/n)\textstyle \sum_{j=1}^{n} H_j(-\delta) \\
		\text{where} \quad H_j(-\delta) &= \frac{U_{1j} \mathit{exp}(\delta U_{1j}) + U_{2j} \mathit{exp}(\delta U_{2j}) - \mathit{exp}(\delta) - (U_{1j} + U_{2j}) \mathit{exp}(\delta(U_{1j} + U_{2j}))}{\mathit{exp}(\delta U_{1j}) + \mathit{exp}(\delta U_{2j}) - \mathit{exp}(\delta) - \mathit{exp}(\delta(U_{1j} + U_{2j}))}.
	\end{align*}
We are going to consider two subcases as follows:\\

	\underline{Subcase - (C1)}: $U_{1j} + U_{2j} > 1$. In this case multiply both the numerator and the denominator of $H_j(-\delta)$ by $\mathit{exp}(-\delta(U_{1j} + U_{2j}))$, and then take the limit as $\delta \to \infty$. It can be seen that $H_j(-\delta) \to (U_{1j} + U_{2j})$.\\
	
	\underline{Subcase - (C2)}:  $U_{1j} + U_{2j} < 1$. In this case multiply both the numerator and the denominator of $H_j(-\delta)$ by $\mathit{exp}(-\delta)$, and then take the limit as $\delta \to \infty$. It can be seen that $H_j(-\delta) \to 1$.\\
	
	Combining the above two subcases it can be seen that as $\delta \to \infty$, each $H_j(-\delta)$ approaches to $\mathit{max}\{(U_{1j} + U_{2j}), 1\} > 0$. Hence,
	\[
	H(-\delta) \to -1 + (\overline{U_1} + \overline{U_2}) + (2/n) \textstyle \sum_{j=1}^{n} \mathit{max}\{(U_{1j} + U_{2j}), 1\} > 0.
	\]
	The above last argument why $ \lim_{\delta \to \infty} H(-\delta) > 0 $ can be seen from the fact that	
	\begin{align*}
		H_j(-\delta) & = -1 - \left(\overline{U}_1 + \overline{U}_2\right) + (2/n) \textstyle \sum_{j=1}^{n} \mathit{max}\{(U_{1j} + U_{2j}), 1\} \\
		\hspace*{2cm} & = (1/n) \textstyle \sum_{j=1}^{n} \left\{-1 + a_j + 2\mathit{max}(a_j, 1)\right\}, \quad a_j = \left(U_{1j} + U_{2j}\right);
	\end{align*}
which is $>0$, since each $a_j = \left(U_{1j} + U_{2j}\right) \in (0,2) $. The above behavior of the function $H(\theta)$ near $-\infty, 0$ and $+\infty$ shows that $H(\theta) = 0$ is guaranteed to have a solution. In other words, $\hat{\theta}_{ML}$ exists. \\
\\
\noindent \textbf{Remark 2.1} To prove that the MLE of $\theta$ obtained by solving the equation (2.4) is indeed unique one must show that the function $H(\theta)$ (the left hand side of (2.4)) is monotonically decreasing, i.e., $H^{'}(\theta)\leq 0, \theta \in \Theta = \mathbb{R} \setminus \{0\}$. But so far we haven't been able to prove it analytically even though it appears to be the case. Therefore, we leave the uniqueness of the solution of (2.4) as a conjecture.
\section{Method of Moment Estimation (MME) of $\theta$}
	When it comes to estimation of $\theta$, two widely used estimators have been proposed by Genest($1987$) \cite{Genest1987}, - one based on the grade correlation coefficient,\\ 
	\begin{equation}
		\rho(\theta) = 12 \int_0^1 \int_0^1 (u_1 - 0.5)(u_2 - 0.5) c(u_1,u_2|\theta) \, du_1 \, du_2; \tag{3.1}
	\end{equation}
and the other one is based on the Kendall's rank correlation coefficient 
	
	\begin{equation}
		\tau(\theta) = 4 \int_0^1 \int_0^1 C(u_1, u_2 | \theta)  c(u_1, u_2 | \theta) \, du_1 du_2 - 1. 	\tag{3.2}
	\end{equation}
[Note that Genest's (1987) \cite{Genest1987} parameter $\alpha$ is our $\mathit{exp}(-\theta)$, i.e., $\theta = \theta(\alpha) = - \mathit{ln}\alpha.$]. Following Genest's(1987) \cite{Genest1987} definition of Debye function $D_k^*(\theta)$, defined as
	\begin{equation}
		D_k^*(\theta) = k\theta^{-k} \int_0^{\theta} \left\{ t^k/(e^t - 1) \right\} dt, \tag{3.3}
	\end{equation}
the above $\tau (\theta)$ and $\rho(\theta)$ can be rewritten as (see p-550, Section 2, Genest(1987) \cite{Genest1987})
	\[
		\tau(\theta) = 1 - (4/\theta)(1 - D_1^*(\theta)), \tag{3.4}
	\]
	
	\[
	\rho(\theta) = 1 - (12/\theta)(D_1^*(\theta) - D_2^*(\theta)). \tag{3.5}
	\]
	
Based on the data $\bm{U}_j = (U_{1j}, U_{2j})',\quad 1 \leq j \leq n$, let $\hat{\tau}$ and $\hat {\rho}$ be the sample analogues of $\tau$ and $\rho$, respectively. In other words, 
	
	\begin{equation}
	\hat{\tau} = \sum_{k=1}^{n-1} \sum_{l=k+1}^{n} U_1^*(k,l) U_2^*(k,l)/\binom{n}{2}, \tag{3.6}
	\end{equation}
where 
	\begin{equation}
	U_i^*(k,l) = 
	\begin{cases} 
	1 & \text{if   }  U_{ik} \leq U_{il}, \\
	-1 & \text{if   }  U_{ik} > U_{il},
	\end{cases} \quad i = 1,2 ;   \tag{3.7}
	\end{equation}
and 
	\begin{equation}
	\hat{\rho} = 1 - 6 \sum_{j=1}^{n} D_j^2 / n(n^2 - 1), \tag{3.8}
	\end{equation}
where 
	\begin{equation}
	D_j = \{ \mathit{rank}(U_{1j}) - \mathit{rank}(U_{2j}) \}, \quad j = 1, 2, \ldots, n; \tag{3.9}
	\end{equation}
(see Genest(1987) \cite{Genest1987} for the details).

Therefore, two moment estimates of $\theta$ can be obtained by replacing $\tau$ and $\rho$ by $\hat{\tau}$ and $\hat{\rho}$ respectively on the LHS of $(3.4)$ and $(3.5)$, and then solve for $\theta$. 
So, the two method of moment estimators of $\theta$ are $\hat{\theta}_{MM1}$ and $\hat{\theta}_{MM2}$ found by solving the following equations, respectively, 
	\[
		1-(4/\theta) \left(1-D_1^*(\theta)\right) = \hat{\tau} \quad \text{(in (3.6)); and}  \tag{3.10}
	\] 
	\[
	1 + (12/\theta)(D_1^*(\theta) - D_2^*(\theta)) = \hat {\rho} \text{ (in (3.8)).}  \tag{3.11}
	\]

\section{Comparison of the Three Estimators of $\theta$}
	In this section we undertake a comprehensive simulation study to compare $\hat{\theta}_{ML}, \hat{\theta}_{MM1}$ and $\hat{\theta}_{MM2}$  in terms of bias, mean squared error(MSE), relative bias and relative MSE.
	
	Generate a random sample $\bm{U} = (U_{1j}, U_{2j}), 1 \leq j \leq n $, of size $n$ from $SBFCD(\theta)$ (as described in subsection 1.4) for a given $\theta \in \Theta = \mathbb{R} \setminus \{0\}$. Compute $\hat{\theta}_{ML}, \hat{\theta}_{MM1}$ and $\hat{\theta}_{MM2}$ to obtain the error in estimation as 
	\[
	e_{ML} = (\hat{\theta}_{ML} - \theta), e_{MM1} = (\hat{\theta}_{MM1} - \theta) \text{ and } e_{MM2} = (\hat{\theta}_{MM2} - \theta).
	\]
	Repeat the above computations a large number of, say $L$, times thereby getting $L$ copies of $e_{ML}, e_{MM1} \text{ and } e_{MM2}$ as $e^{(l)}_{ML}, e^{(l)}_{MM1} \text{ and } e^{(l)}_{MM2}, 1 \leq l \leq L $. The bias and MSE are then approximated by 
	\[
	\text{Bias}(\hat{\theta}_\ast) = \textstyle \sum_{l=1}^{L} e_\ast^{(l)}/L \quad , \quad 
	\text{MSE}(\hat{\theta}_\ast) = \textstyle \sum_{l=1}^{L} (e_\ast^{(l)})^2/L
	\]
where the subscript `*' represents any of the three estimators (ML, MM1, MM2).
The following Tables 4.1(a) - (b) shows the bias (Bias) and MSE values of the three estimators for selected $n$ and $\theta$ based on $L= 20,000$ replications.

Tables 4.1(c) - (d) present relative bias (or, RBias) = Bias/$|\theta|$, and relative MSE  (or, RMSE) = MSE$/\theta^2$ of the three above estimators.\\
\\
\noindent \textbf{Remark 4.1.} The observed simulation results have some interesting patterns and symmetry, and hence, they have been presented for $\theta > 0$ only. It has been noted that for $\theta > 0$, 
\begin{enumerate}[label=(\alph*)] 
	\item Bias of the MLE at $(-\theta)=-$(Bias of the MLE at $\theta$), i.e., the Bias of the MLE is an odd function of $\theta$. For $\theta > 0$, the MLE over estimates $\theta$, and it under estimates when $\theta < 0$. Also, the magnitude of bias increases as $\theta$ moves aways from $0$.
	\item  Bias of each MME acts in the opposite direction compared to that of the MLE. Bias of each MME at $(-\theta)$ = $ - $(Bias of the MME at $\theta) > 0$. Each MME overestimates the parameter $\theta$  when it is negative, and underestimates it when $\theta$ is positive.
	\item In terms of bias magnitude, the MLE appears to be far better than the MMEs. All the three point estimators appear to have zero bias when $\theta = 0$.
	\item Both the MMEs have near identical performance both in terms of bias as well as MSE.
	\item In terms of MSE, the MLE again tends to perform better than the MMEs for all values of $\theta$ except for a small neighborhood around $0$.
	\item RBias and RMSE values show the relative deviance of each estimator with respect to the value of $\theta$. This also shows that near $\theta = 0$, especially for a small sample size, the MLE can be dominated by the MMEs.
\end{enumerate}
\setcounter{table}{0}
\renewcommand{\thetable}{4.1(\alph{table})}
\begin{table}[H]
	\centering
	\caption{Simulated Bias and MSE of three estimatiors of $\theta$ $(n=5,10,15,20)$}
	\renewcommand{\arraystretch}{0.9} 
	\renewcommand{\baselinestretch}{0.9} 
	\small 
	\resizebox{1.0\textwidth}{!}{  
		\begin{tabular}{|c|c|c|c|c||c|c|c|}
			\hline
			$n$ & $\theta$ &  Bias-MLE & Bias-MME1 & Bias-MME2 & MSE-MLE & MSE-MME1 & MSE-MME2\\
			\hline
			\multirow{14}{*}{5} & 10 & 2.049 & -7.467 & -7.866 & 57.580 & 83.919 & 78.007 \\
			& 9  & 1.848 & -6.505 & -6.940 & 51.288 & 69.038 & 63.406 \\
			& 8  & 1.659 & -5.585 & -5.946 & 44.921 & 56.412 & 50.279 \\
			& 7  & 1.422 & -4.688 & -5.008 & 36.228 & 46.206 & 39.239 \\
			& 6  & 1.323 & -3.829 & -4.138 & 31.497 & 36.506 & 30.277 \\
			& 5  & 1.070 & -2.997 & -3.333 & 26.167 & 29.232 & 22.836 \\
			& 4  & 0.769 & -2.296 & -2.575 & 21.301 & 22.976 & 17.134 \\
			& 3  & 0.602 & -1.634 & -1.844 & 18.755 & 17.831 & 13.003 \\
			& 2  & 0.403 & -1.053 & -1.202 & 17.003 & 14.066 & 9.707 \\
			& 1.5& 0.279 & -0.807 & -0.883 & 15.533 & 13.131 & 8.841 \\
			& 1  & 0.200 & -0.514 & -0.578 & 15.496 & 12.323 & 8.249 \\
			& 0.75& 0.130 & -0.385 & -0.411 & 15.240 & 12.107 & 7.963 \\
			& 0.5 & 0.099 & -0.265 & -0.271 & 15.165 & 11.494 & 7.795 \\
			& 0.1 & 0.007 & -0.052 & -0.049 & 15.038 & 11.180 & 7.564 \\
			\hline
			\multirow{14}{*}{10}& 10 & 0.992 & -6.696 & -7.133 & 17.659 & 76.353 & 73.830 \\
			& 9  & 0.912 & -5.819 & -6.130 & 15.075 & 62.423 & 58.507 \\
			& 8  & 0.819 & -5.011 & -5.336 & 12.789 & 50.035 & 46.277 \\
			& 7  & 0.788 & -4.227 & -4.551 & 11.590 & 38.555 & 35.925 \\
			& 6  & 0.629 & -3.497 & -3.773 & 9.389 & 28.812 & 26.875 \\
			& 5  & 0.529 & -2.828 & -3.054 & 8.033 & 20.455 & 19.015 \\
			& 4  & 0.422 & -2.157 & -2.369 & 7.196 & 14.225 & 13.113 \\
			& 3  & 0.321 & -1.545 & -1.706 & 6.143 & 9.321 & 8.430 \\
			& 2  & 0.184 & -1.047 & -1.135 & 5.601 & 5.872 & 5.224 \\
			& 1.5& 0.144 & -0.768 & -0.818 & 5.511 & 4.629 & 4.079 \\
			& 1  & 0.104 & -0.512 & -0.542 & 5.399 & 3.750 & 3.216 \\
			& 0.75& 0.082 & -0.378 & -0.420 & 5.318 & 3.497 & 2.958 \\
			& 0.5 & 0.074 & -0.227 & -0.245 & 5.206 & 3.317 & 2.693 \\
			& 0.1 & 0.028 & -0.046 & -0.044 & 5.169 & 3.007 & 2.418 \\
			\hline
			\multirow{14}{*}{15} & 10  & 0.643 & -6.529 & -6.952 & 9.865 & 72.437 & 71.096 \\
			& 9   & 0.569 & -5.663 & -6.119 & 8.379 & 57.658 & 56.866 \\
			& 8   & 0.516 & -4.969 & -5.295 & 7.152 & 45.606 & 44.618 \\
			& 7   & 0.455 & -4.159 & -4.487 & 6.177 & 34.501 & 33.925 \\
			& 6   & 0.436 & -3.486 & -3.720 & 5.344 & 25.477 & 24.943 \\
			& 5   & 0.387 & -2.817 & -3.022 & 4.683 & 17.950 & 17.418 \\
			& 4   & 0.289 & -2.212 & -2.360 & 4.057 & 11.997 & 11.638 \\
			& 3   & 0.212 & -1.614 & -1.723 & 3.517 & 7.188  & 7.057  \\
			& 2   & 0.140 & -1.049 & -1.125 & 3.350 & 4.112  & 3.955  \\
			& 1.5 & 0.103 & -0.787 & -0.852 & 3.210 & 3.008  & 2.848  \\
			& 1   & 0.083 & -0.521 & -0.560 & 3.191 & 2.258  & 2.085  \\
			& 0.75& 0.048 & -0.392 & -0.425 & 3.178 & 2.044  & 1.825  \\
			& 0.5 & 0.022 & -0.261 & -0.285 & 3.157 & 1.816  & 1.611  \\
			& 0.1 & 0.008 & -0.048 & -0.057 & 3.146 & 1.671  & 1.472  \\
			\hline
			\multirow{14}{*}{20} & 10  & 0.485 & -6.447 & -6.904 & 6.805 & 69.659 & 69.960 \\
			& 9   & 0.444 & -5.659 & -6.045 & 5.904 & 55.741 & 55.906 \\
			& 8   & 0.415 & -4.972 & -5.235 & 5.189 & 43.839 & 43.682 \\
			& 7   & 0.353 & -4.197 & -4.465 & 4.398 & 33.007 & 33.008 \\
			& 6   & 0.312 & -3.501 & -3.689 & 3.683 & 24.120 & 23.976 \\
			& 5   & 0.274 & -2.843 & -3.026 & 3.223 & 16.733 & 16.821 \\
			& 4   & 0.213 & -2.215 & -2.360 & 2.769 & 10.852 & 10.968 \\
			& 3   & 0.165 & -1.638 & -1.741 & 2.506 & 6.530  & 6.593  \\
			& 2   & 0.122 & -1.059 & -1.144 & 2.367 & 3.523  & 3.475  \\
			& 1.5 & 0.085 & -0.785 & -0.846 & 2.244 & 2.449  & 2.405  \\
			& 1   & 0.044 & -0.540 & -0.557 & 2.210 & 1.706  & 1.662  \\
			& 0.75& 0.039 & -0.398 & -0.428 & 2.191 & 1.441  & 1.363  \\
			& 0.5 & 0.016 & -0.279 & -0.289 & 2.181 & 1.240  & 1.175  \\
			& 0.1 & 0.001 & -0.058 & -0.053 & 2.174 & 1.148  & 1.059  \\
			\hline
		\end{tabular}
	}
\end{table}

\begin{table}[H]
	\centering
	\caption{Simulated Bias and MSE of three estimatiors of $\theta$ $(n=25,50,75,100)$}
	\renewcommand{\arraystretch}{0.95} 
	\renewcommand{\baselinestretch}{0.9} 
	\small 
	\resizebox{1.0\textwidth}{!}{  
		\begin{tabular}{|c|c|c|c|c||c|c|c|}
			\hline
			$n$ & $\theta$ & Bias-MLE & Bias-MME1 & Bias-MME2 & MSE-MLE & MSE-MME1 & MSE-MME2 \\
			\hline
		\multirow{14}{*}{25} & 10  & 0.385 & -6.456 & -6.845 & 5.287 & 68.225 & 69.512 \\
		& 9   & 0.342 & -5.701 & -6.040 & 4.524 & 54.792 & 55.416 \\
		& 8   & 0.300 & -4.991 & -5.297 & 3.885 & 42.965 & 43.489 \\
		& 7   & 0.287 & -4.203 & -4.440 & 3.381 & 32.128 & 32.516 \\
		& 6   & 0.243 & -3.496 & -3.694 & 2.890 & 23.310 & 23.565 \\
		& 5   & 0.224 & -2.845 & -3.004 & 2.493 & 16.104 & 16.393 \\
		& 4   & 0.183 & -2.239 & -2.347 & 2.170 & 10.491 & 10.678 \\
		& 3   & 0.113 & -1.634 & -1.729 & 1.926 & 6.127  & 6.269  \\
		& 2   & 0.088 & -1.070 & -1.124 & 1.788 & 3.186  & 3.241  \\
		& 1.5 & 0.060 & -0.797 & -0.846 & 1.745 & 2.170  & 2.173  \\
		& 1   & 0.045 & -0.534 & -0.568 & 1.729 & 1.449  & 1.428  \\
		& 0.75& 0.032 & -0.397 & -0.434 & 1.720 & 1.180  & 1.153  \\
		& 0.5 & 0.026 & -0.270 & -0.283 & 1.715 & 0.991  & 0.961  \\
		& 0.1 & 0.012 & -0.050 & -0.049 & 1.705 & 0.873  & 0.806  \\
		\hline
		\multirow{14}{*}{50} & 10  & 0.181 & -6.415 & -6.759 & 2.381 & 65.890 & 68.084 \\
		& 9   & 0.175 & -5.725 & -5.996 & 2.083 & 53.083 & 54.483 \\
		& 8   & 0.154 & -4.946 & -5.174 & 1.775 & 40.975 & 42.031 \\
		& 7   & 0.152 & -4.232 & -4.413 & 1.571 & 30.885 & 31.720 \\
		& 6   & 0.127 & -3.547 & -3.650 & 1.340 & 22.306 & 22.743 \\
		& 5   & 0.112 & -2.886 & -3.009 & 1.159 & 15.327 & 15.760 \\
		& 4   & 0.088 & -2.258 & -2.339 & 1.000 & 9.712  & 10.047 \\
		& 3   & 0.073 & -1.647 & -1.723 & 0.910 & 5.528  & 5.761  \\
		& 2   & 0.041 & -1.098 & -1.144 & 0.831 & 2.679  & 2.805  \\
		& 1.5 & 0.034 & -0.814 & -0.850 & 0.808 & 1.687  & 1.763  \\
		& 1   & 0.030 & -0.543 & -0.576 & 0.806 & 0.968  & 1.037  \\
		& 0.75& 0.023 & -0.396 & -0.430 & 0.789 & 0.718  & 0.752  \\
		& 0.5 & 0.013 & -0.272 & -0.291 & 0.785 & 0.546  & 0.552  \\
		& 0.1 & 0.008 & -0.050 & -0.056 & 0.783 & 0.414  & 0.411  \\
		\hline
		\multirow{14}{*}{75} & 10  & 0.114 & -6.503 & -6.797 & 1.534 & 66.105 & 68.259 \\
		& 9   & 0.107 & -5.692 & -5.994 & 1.351 & 52.191 & 54.416 \\
		& 8   & 0.104 & -4.993 & -5.202 & 1.168 & 40.807 & 42.122 \\
		& 7   & 0.093 & -4.183 & -4.451 & 1.005 & 30.087 & 31.663 \\
		& 6   & 0.075 & -3.572 & -3.711 & 0.859 & 22.062 & 22.833 \\
		& 5   & 0.067 & -2.896 & -3.005 & 0.738 & 15.047 & 15.544 \\
		& 4   & 0.060 & -2.263 & -2.326 & 0.658 & 9.518  & 9.779  \\
		& 3   & 0.053 & -1.645 & -1.715 & 0.582 & 5.336  & 5.593  \\
		& 2   & 0.038 & -1.090 & -1.137 & 0.546 & 2.525  & 2.658  \\
		& 1.5 & 0.026 & -0.819 & -0.869 & 0.520 & 1.533  & 1.657  \\
		& 1   & 0.013 & -0.540 & -0.575 & 0.515 & 0.831  & 0.900  \\
		& 0.75& 0.011 & -0.408 & -0.440 & 0.510 & 0.585  & 0.629  \\
		& 0.5 & 0.004 & -0.277 & -0.298 & 0.509 & 0.410  & 0.438  \\
		& 0.1 & 0.001 & -0.059 & -0.058 & 0.508 & 0.275  & 0.281  \\
		\hline
		\multirow{14}{*}{100} & 10  & 0.091 & -6.514 & -6.681 & 1.165 & 66.015 & 67.167 \\
		& 9   & 0.083 & -5.693 & -5.971 & 0.996 & 52.017 & 54.098 \\
		& 8   & 0.080 & -4.960 & -5.153 & 0.876 & 40.391 & 41.632 \\
		& 7   & 0.062 & -4.233 & -4.403 & 0.733 & 30.199 & 31.284 \\
		& 6   & 0.061 & -3.570 & -3.667 & 0.643 & 21.929 & 22.519 \\
		& 5   & 0.054 & -2.899 & -2.968 & 0.547 & 14.947 & 15.298 \\
		& 4   & 0.042 & -2.248 & -2.316 & 0.488 & 9.364  & 9.682  \\
		& 3   & 0.030 & -1.653 & -1.717 & 0.430 & 5.255  & 5.513  \\
		& 2   & 0.025 & -1.083 & -1.141 & 0.398 & 2.435  & 2.600  \\
		& 1.5 & 0.023 & -0.817 & -0.871 & 0.389 & 1.472  & 1.602  \\
		& 1   & 0.014 & -0.542 & -0.591 & 0.384 & 0.778  & 0.854  \\
		& 0.75& 0.007 & -0.412 & -0.443 & 0.377 & 0.524  & 0.579  \\
		& 0.5 & 0.004 & -0.277 & -0.302 & 0.374 & 0.347  & 0.376  \\
		& 0.1 & 0.001 & -0.060 & -0.054 & 0.365 & 0.202  & 0.219  \\
		\hline
		
	\end{tabular}
}
\end{table}

\begin{table}[H]
	\centering
	\caption{Simulated RBias and RMSE of three estimatiors of $\theta$ $(n=5,10,15,20)$}
	\renewcommand{\arraystretch}{0.95} 
	\renewcommand{\baselinestretch}{0.97} 
	\small 
	\resizebox{1.0\textwidth}{!}{  
		\begin{tabular}{|c|c|c|c|c||c|c|c|c|}
			\hline
			$n$ & $\theta$ & RBias-MLE & RBias-MME1 & RBias-MME2 & RMSE-MLE & RMSE-MME1 & RMSE-MME2 \\
			\hline
			\multirow{14}{*}{5}  & 10  & 0.205 & -0.747 & -0.787 & 0.576 & 0.839 & 0.780 \\
			& 9   & 0.205 & -0.723 & -0.771 & 0.633 & 0.852 & 0.783 \\
			& 8   & 0.207 & -0.698 & -0.743 & 0.702 & 0.881 & 0.786 \\
			& 7   & 0.203 & -0.670 & -0.715 & 0.739 & 0.943 & 0.801 \\
			& 6   & 0.220 & -0.638 & -0.690 & 0.875 & 1.014 & 0.841 \\
			& 5   & 0.214 & -0.599 & -0.667 & 1.047 & 1.169 & 0.913 \\
			& 4   & 0.192 & -0.574 & -0.644 & 1.331 & 1.436 & 1.071 \\
			& 3   & 0.201 & -0.545 & -0.615 & 2.084 & 1.981 & 1.445 \\
			& 2   & 0.202 & -0.527 & -0.601 & 4.251 & 3.516 & 2.427 \\
			& 1.5 & 0.186 & -0.538 & -0.589 & 6.904 & 5.836 & 3.930 \\
			& 1   & 0.200 & -0.514 & -0.578 & 15.496 & 12.323 & 8.249 \\
			& 0.75& 0.173 & -0.513 & -0.548 & 27.094 & 21.523 & 14.156\\
			& 0.5 & 0.199 & -0.531 & -0.541 & 60.658 & 45.976 & 31.181 \\
			& 0.1 & 0.073 & -0.518 & -0.486 & 1503.765 & 1117.973 & 756.448 \\
			\hline
			\multirow{14}{*}{10} & 10  & 0.099 & -0.670 & -0.713 & 0.177 & 0.764 & 0.738 \\
			& 9   & 0.101 & -0.647 & -0.681 & 0.186 & 0.771 & 0.722 \\
			& 8   & 0.102 & -0.626 & -0.667 & 0.200 & 0.782 & 0.723 \\
			& 7   & 0.113 & -0.604 & -0.650 & 0.237 & 0.787 & 0.733 \\
			& 6   & 0.105 & -0.583 & -0.629 & 0.261 & 0.800 & 0.747 \\
			& 5   & 0.106 & -0.566 & -0.611 & 0.321 & 0.818 & 0.761 \\
			& 4   & 0.105 & -0.539 & -0.592 & 0.450 & 0.889 & 0.820 \\
			& 3   & 0.107 & -0.515 & -0.569 & 0.683 & 1.036 & 0.937 \\
			& 2   & 0.092 & -0.523 & -0.568 & 1.400 & 1.468 & 1.306 \\
			& 1.5 & 0.096 & -0.512 & -0.545 & 2.449 & 2.057 & 1.813 \\
			& 1   & 0.104 & -0.512 & -0.542 & 5.399 & 3.750 & 3.216 \\
			& 0.75& 0.109 & -0.504 & -0.559 & 9.454 & 6.218 & 5.259 \\
			& 0.5 & 0.149 & -0.453 & -0.490 & 20.822 & 13.270 & 10.772 \\
			& 0.1 & 0.277 & -0.460 & -0.441 & 516.858 & 300.741 & 241.778 \\
			\hline
			\multirow{14}{*}{15} 
						
			&10    & 0.064 & -0.653 & -0.695 & 0.099  & 0.724   & 0.711   \\
			&9     & 0.063 & -0.629 & -0.680 & 0.103  & 0.712   & 0.702   \\
			&8     & 0.064 & -0.621 & -0.662 & 0.112  & 0.713   & 0.697   \\
			&7     & 0.065 & -0.594 & -0.641 & 0.126  & 0.704   & 0.692   \\
			&6     & 0.073 & -0.581 & -0.620 & 0.148  & 0.708   & 0.693   \\
			&5     & 0.077 & -0.563 & -0.604 & 0.187  & 0.718   & 0.697   \\
			&4     & 0.072 & -0.553 & -0.590 & 0.254  & 0.750   & 0.727   \\
			&3     & 0.071 & -0.538 & -0.574 & 0.391  & 0.799   & 0.784   \\
			&2     & 0.070 & -0.525 & -0.563 & 0.838  & 1.028   & 0.989   \\
			&1.5   & 0.069 & -0.525 & -0.568 & 1.427  & 1.337   & 1.266   \\
			&1     & 0.083 & -0.521 & -0.560 & 3.191  & 2.258   & 2.085   \\
			&0.75  & 0.064 & -0.522 & -0.566 & 5.650  & 3.634   & 3.244   \\
			&0.5   & 0.043 & -0.522 & -0.570 & 12.627 & 7.263   & 6.443   \\
			&0.1   & 0.079 & -0.483 & -0.566 & 314.563 & 167.068 & 147.245 \\
			\hline
			\multirow{14}{*}{20} 
			& 10    & 0.049 & -0.645 & -0.690 & 0.068  & 0.697   & 0.700   \\
			&9     & 0.049 & -0.629 & -0.672 & 0.073  & 0.688   & 0.690   \\
			&8     & 0.052 & -0.621 & -0.654 & 0.081  & 0.685   & 0.683   \\
			&7     & 0.050 & -0.600 & -0.638 & 0.090  & 0.674   & 0.674   \\
			&6     & 0.052 & -0.584 & -0.615 & 0.102  & 0.670   & 0.666   \\
			&5     & 0.055 & -0.569 & -0.605 & 0.129  & 0.669   & 0.673   \\
			&4     & 0.053 & -0.554 & -0.590 & 0.173  & 0.678   & 0.686   \\
			&3     & 0.055 & -0.546 & -0.580 & 0.278  & 0.726   & 0.733   \\
			&2     & 0.061 & -0.529 & -0.572 & 0.592  & 0.881   & 0.869   \\
			&1.5   & 0.057 & -0.524 & -0.564 & 0.997  & 1.088   & 1.069   \\
			&1     & 0.044 & -0.540 & -0.557 & 2.210  & 1.706   & 1.662   \\
			&0.75  & 0.052 & -0.531 & -0.571 & 3.896  & 2.562   & 2.423   \\
			&0.5   & 0.031 & -0.558 & -0.577 & 8.723  & 4.960   & 4.701   \\
			&0.1   & 0.008 & -0.582 & -0.535 & 217.359 & 114.842 & 105.940 \\
			\hline
		\end{tabular}
	}
\end{table}

\begin{table}[H]
	\centering
	\caption{Simulated RBias and RMSE of three estimators of $\theta$ $(n=25, 50, 75, 100)$}
	\renewcommand{\arraystretch}{0.95} 
	\small 
	\resizebox{1.0\textwidth}{!}{  
		\begin{tabular}{|c|c|c|c|c||c|c|c|c|}
			\hline
			$n$ & $\theta$ & RBias-MLE & RBias-MME1 & RBias-MME2 & RMSE-MLE & RMSE-MME1 & RMSE-MME2 \\
			\hline
			\multirow{14}{*}{25} 
			& 10    & 0.039 & -0.646 & -0.685 & 0.053  & 0.682   & 0.695   \\
			&9     & 0.038 & -0.633 & -0.671 & 0.056  & 0.676   & 0.684   \\
			&8     & 0.037 & -0.624 & -0.662 & 0.061  & 0.671   & 0.680   \\
			&7     & 0.041 & -0.600 & -0.634 & 0.069  & 0.656   & 0.664   \\
			&6     & 0.041 & -0.583 & -0.616 & 0.080  & 0.648   & 0.655   \\
			&5     & 0.045 & -0.569 & -0.601 & 0.100  & 0.644   & 0.656   \\
			&4     & 0.046 & -0.560 & -0.587 & 0.136  & 0.656   & 0.667   \\
			&3     & 0.038 & -0.545 & -0.576 & 0.214  & 0.681   & 0.697   \\
			&2     & 0.044 & -0.535 & -0.562 & 0.447  & 0.796   & 0.810   \\
			&1.5   & 0.040 & -0.532 & -0.564 & 0.776  & 0.964   & 0.966   \\
			&1     & 0.045 & -0.534 & -0.568 & 1.729  & 1.449   & 1.428   \\
			&0.75  & 0.042 & -0.529 & -0.578 & 3.058  & 2.098   & 2.049   \\
			&0.5   & 0.052 & -0.540 & -0.565 & 6.860  & 3.963   & 3.846   \\
			&0.1   & 0.125 & -0.499 & -0.493 & 170.494 & 87.308 & 80.612  \\
			\hline
			\multirow{14}{*}{50} 
			& 10    & 0.018 & -0.642 & -0.676 & 0.024  & 0.659   & 0.681   \\
			&9     & 0.019 & -0.636 & -0.666 & 0.026  & 0.655   & 0.673   \\
			&8     & 0.019 & -0.618 & -0.647 & 0.028  & 0.640   & 0.657   \\
			&7     & 0.022 & -0.605 & -0.630 & 0.032  & 0.630   & 0.647   \\
			&6     & 0.021 & -0.591 & -0.608 & 0.037  & 0.620   & 0.632   \\
			&5     & 0.022 & -0.577 & -0.602 & 0.046  & 0.613   & 0.630   \\
			&4     & 0.022 & -0.564 & -0.585 & 0.063  & 0.607   & 0.628   \\
			&3     & 0.024 & -0.549 & -0.574 & 0.101  & 0.614   & 0.640   \\
			&2     & 0.021 & -0.549 & -0.572 & 0.208  & 0.670   & 0.701   \\
			&1.5   & 0.023 & -0.543 & -0.567 & 0.359  & 0.750   & 0.783   \\
			&1     & 0.030 & -0.543 & -0.576 & 0.806  & 0.968   & 1.037   \\
			&0.75  & 0.031 & -0.528 & -0.573 & 1.403  & 1.276   & 1.337   \\
			&0.5   & 0.025 & -0.544 & -0.583 & 3.142  & 2.185   & 2.209   \\
			&0.1   & 0.079 & -0.501 & -0.562 & 78.338 & 41.404  & 41.116  \\
			
			\hline
			\multirow{14}{*}{75} 
			&10    & 0.011 & -0.650 & -0.680 & 0.015  & 0.661   & 0.683   \\
			&9     & 0.012 & -0.632 & -0.666 & 0.017  & 0.644   & 0.672   \\
			&8     & 0.013 & -0.624 & -0.650 & 0.018  & 0.638   & 0.658   \\
			&7     & 0.013 & -0.598 & -0.636 & 0.021  & 0.614   & 0.646   \\
			&6     & 0.012 & -0.595 & -0.619 & 0.024  & 0.613   & 0.634   \\
			&5     & 0.013 & -0.579 & -0.601 & 0.030  & 0.602   & 0.622   \\
			&4     & 0.015 & -0.566 & -0.582 & 0.041  & 0.595   & 0.611   \\
			&3     & 0.018 & -0.548 & -0.572 & 0.065  & 0.593   & 0.621   \\
			&2     & 0.019 & -0.545 & -0.568 & 0.136  & 0.631   & 0.665   \\
			&1.5   & 0.017 & -0.546 & -0.579 & 0.231  & 0.681   & 0.736   \\
			&1     & 0.013 & -0.540 & -0.575 & 0.515  & 0.831   & 0.900   \\
			&0.75  & 0.015 & -0.543 & -0.587 & 0.906  & 1.040   & 1.118   \\
			&0.5   & 0.008 & -0.555 & -0.597 & 2.034  & 1.640   & 1.750   \\
			&0.1   & 0.014 & -0.593 & -0.584 & 50.753 & 27.512  & 28.128  \\
			\hline
			\multirow{14}{*}{100} 
			
			& 10    & 0.009 & -0.651 & -0.668 & 0.012  & 0.660   & 0.672   \\
			&9     & 0.009 & -0.633 & -0.663 & 0.012  & 0.642   & 0.668   \\
			&8     & 0.010 & -0.620 & -0.644 & 0.014  & 0.631   & 0.651   \\
			&7     & 0.009 & -0.605 & -0.629 & 0.015  & 0.616   & 0.638   \\
			&6     & 0.010 & -0.595 & -0.611 & 0.018  & 0.609   & 0.626   \\
			&5     & 0.011 & -0.580 & -0.594 & 0.022  & 0.598   & 0.612   \\
			&4     & 0.011 & -0.562 & -0.579 & 0.031  & 0.585   & 0.605   \\
			&3     & 0.010 & -0.551 & -0.572 & 0.048  & 0.584   & 0.613   \\
			&2     & 0.012 & -0.542 & -0.571 & 0.099  & 0.609   & 0.650   \\
			&1.5   & 0.016 & -0.545 & -0.580 & 0.173  & 0.654   & 0.712   \\
			&1     & 0.014 & -0.542 & -0.591 & 0.384  & 0.778   & 0.854   \\
			&0.75  & 0.009 & -0.550 & -0.591 & 0.671  & 0.932   & 1.029   \\
			&0.5   & 0.009 & -0.553 & -0.603 & 1.494  & 1.388   & 1.506   \\
			&0.1   & 0.014 & -0.596 & -0.537 & 36.532 & 20.154  & 21.911  \\
			
			\hline
		\end{tabular}
	}
\end{table}

\begin{figure}[H]
	\centering
	\includegraphics[width=1.0\textwidth]{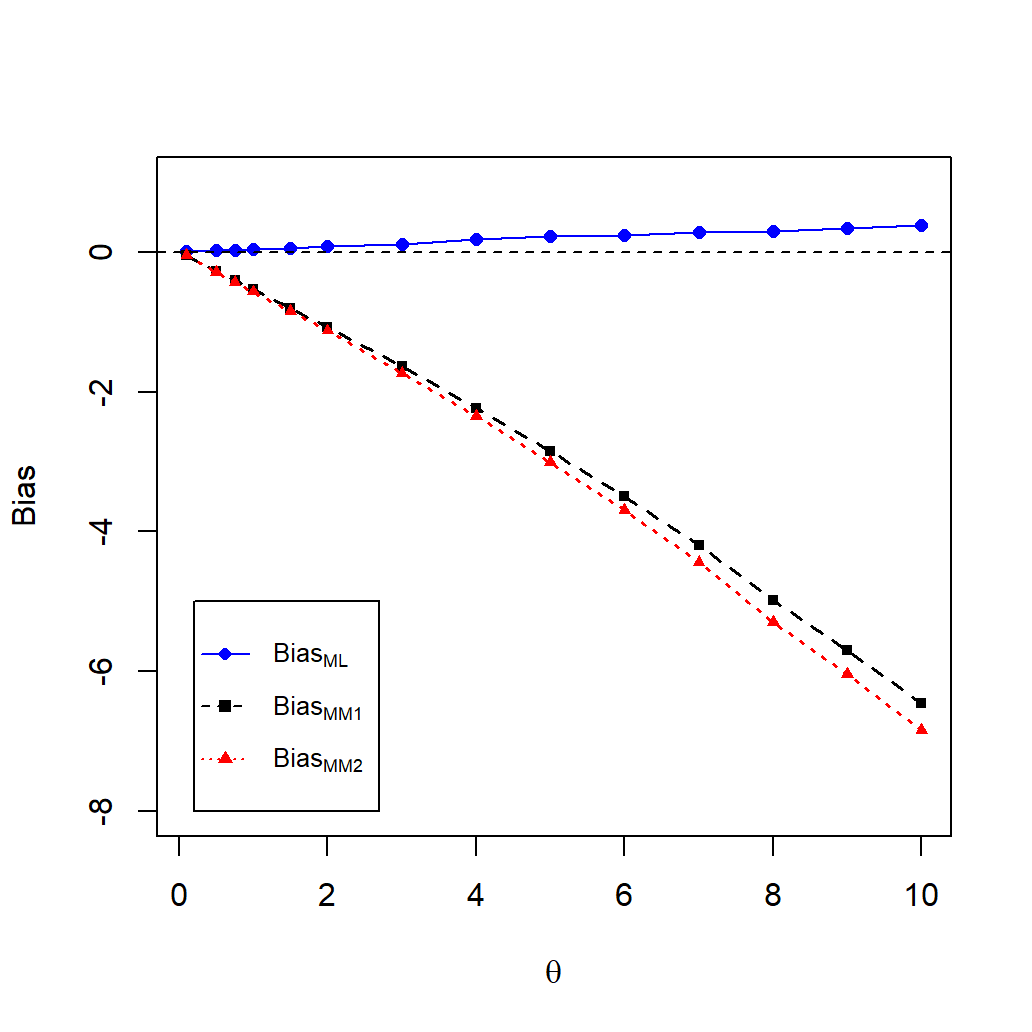} 
	\caption{Figure 4.1(a) Bias plots with $n = 25$}
	\label{fig:a}
\end{figure}

\begin{figure}[H]
	\centering
	\includegraphics[width=1.0\textwidth]{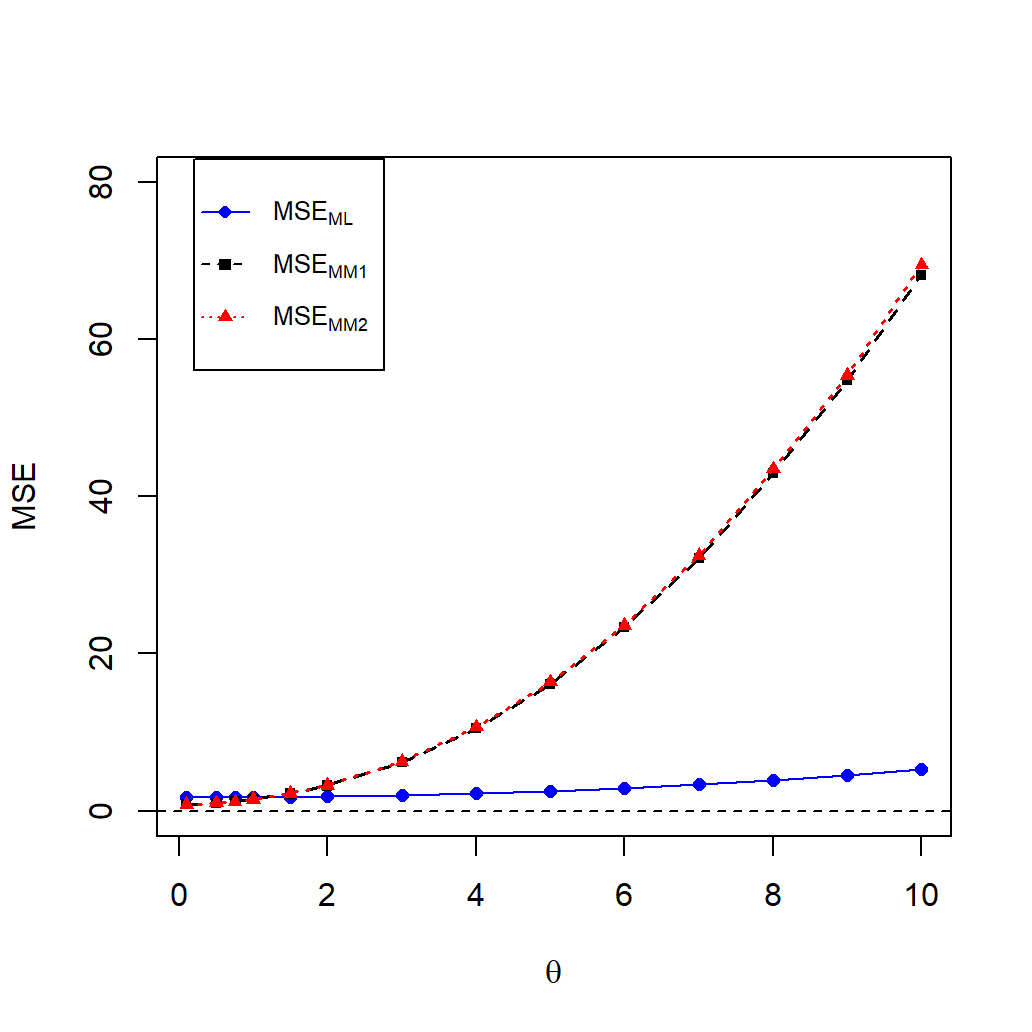} 
	\caption{Figure 4.1(b) MSE plots with $n = 25$}
	\label{fig:a}
\end{figure}

\begin{figure}[H]
	\centering
	\includegraphics[width=1.0\textwidth]{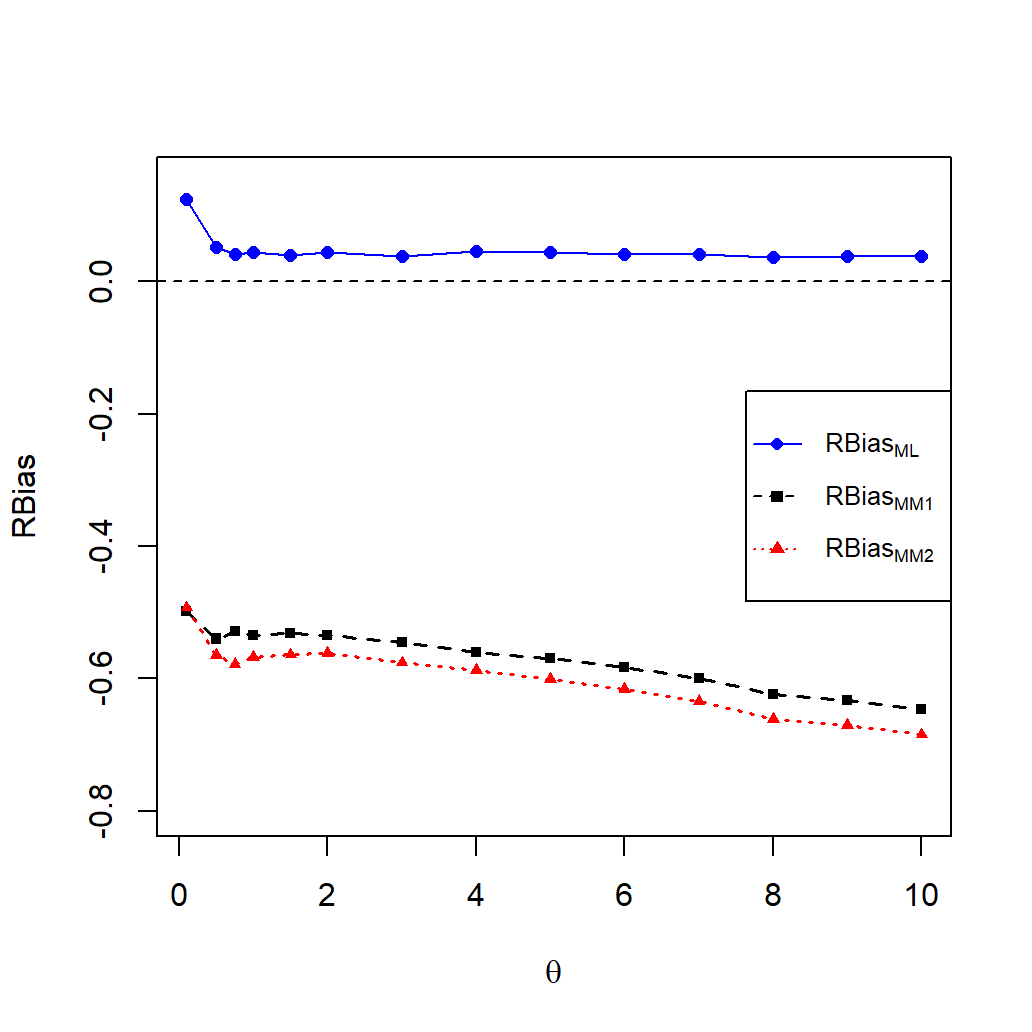} 
	\caption{Figure 4.1(c) RBias plots with $n = 25$}
	\label{fig:a}
\end{figure}

\begin{figure}[H]
	\centering
	\includegraphics[width=1.0\textwidth]{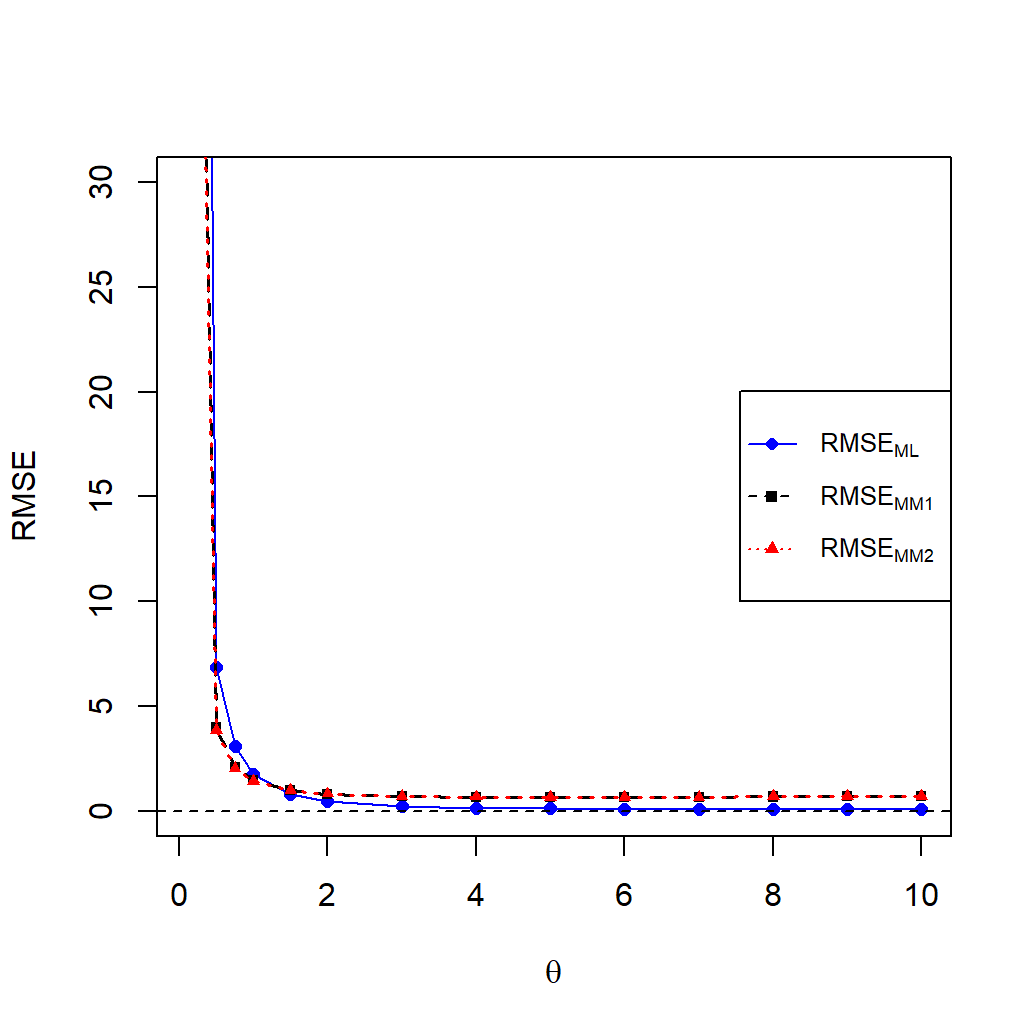} 
	\caption{Figure 4.1(d) RMSE plots with $n = 25$}
	\label{fig:a}
\end{figure}

\noindent \textbf{Remark 4.2.} The standard error (SE) of each simulated bias value can be seen by looking at the corresponding MSE, since $SE\cong~\sqrt{MSE/20,000}$. For example, with $n=25,$ the SE of the MLE's bias range from $0.009 \text{ (at } \theta = 0.1)$ to $0.016 \text{ (at }\theta = 10.0)$. Similarly, the SE of the bias of MME2 is almost identical to that of MME1.\\
\\
\textbf{Remark 4.3.} The SE of the simulated MSE show that (though not fully reported here for brevity) the tabulated values presented in Table 4.1 (a) - (b) are fairly accurate. As an example, the following Table 4.2 shows the SE of simulated MSEs of three estimators for $n=25$. \\
\renewcommand{\thetable}{4.2}

\begin{table}[h!]
	\centering
	\caption{Standard Error (SE) of simulated MSEs ($n = 25$)}
	\label{tab:se_mse}
	\renewcommand{\arraystretch}{1.0} 
	\setlength{\tabcolsep}{9pt} 
	
	\resizebox{0.95\textwidth}{!}{ 
		\begin{tabular}{|c| c c c || c| c c c|}
			\hline
			\textbf{$\theta$} & \textbf{MLE} & \textbf{MME1} & \textbf{MME2} & \textbf{$\theta$} & \textbf{MLE} & \textbf{MME1} & \textbf{MME2} \\
			\hline
			0.1  & 0.020 & 0.015 & 0.017 & 4  & 0.028 & 0.045 & 0.052 \\
			0.5  & 0.023 & 0.015 & 0.013 & 5  & 0.029 & 0.083 & 0.071 \\
			0.75 & 0.015 & 0.011 & 0.012 & 6  & 0.041 & 0.114 & 0.083 \\
			1    & 0.019 & 0.018 & 0.012 & 7  & 0.032 & 0.160 & 0.175 \\
			1.5  & 0.025 & 0.014 & 0.012 & 8  & 0.063 & 0.208 & 0.222 \\
			2    & 0.018 & 0.019 & 0.015 & 9  & 0.059 & 0.262 & 0.199 \\
			3    & 0.024 & 0.028 & 0.033 & 10 & 0.071 & 0.242 & 0.294 \\
			\hline
		\end{tabular}
	}
\end{table}
	\newpage

\noindent \textbf{Remark 4.4.} As mentioned in the previous remark, and evident from Tables 4.1(a)-(d), the MLE and the MMEs have biases acting in opposite directions. While the MLE tends to over (under) estimate for positive (negative) values of $\theta$, both the MMEs do just the opposite. Why the biases are behaving like this is not clear, but this observation is totally new, and to best of our knowledge this wasn't reported before. Also, in terms of absolute bias, the MLE is doing for better then the MMEs. However, in terms of MSE, things are a bit complicated. For $|\theta|$ away from $0$, the MLE is definitely better than the two MMEs; but as $|\theta|$ gets closer to $0$, the MMEs tends to perform (especially MME2) far better than the MLE. This can be gauged by looking at the RMSE values, particularly for ``small" $n$. Also, as expected, the bias and MSE values of all the three estimators decrease monotonically with respect to $n$ for every $\theta$ pointwise.\\
\\
\textbf{Remark 4.5.} Even though the MLE shows much better overall performance than the MMEs, the latters do have lower MSE than the former when $\theta$ lies in a (small) neighborhood of $0$. The following Table 4.3 shows the value of $\varepsilon$ for which at least one MME is better than the MLE in terms of MSE over the interval $ \{|\theta| \leq \varepsilon \}$.

\renewcommand{\thetable}{4.3}
\begin{table}[h!]
	\centering
	\caption{The MME having the lowest MSE (and better than the MLE) over \( |\theta|\leq \varepsilon\)}
	
	\renewcommand{\arraystretch}{1.2} 
	\setlength{\tabcolsep}{9pt} 
	
	\resizebox{0.95\textwidth}{!}{ 
		\begin{tabular}{|c|c|c||c|c|c|}
			\hline
			\( n \) & \( \varepsilon \) & \textbf{MME superior to the MLE} & \( n \) & \( \varepsilon \) & \textbf{MME superior to the MLE} \\
			\hline
			5  & 6.0  & MME2  & 25  & 1.0  & MME2  \\
			10 & 2.0  & MME2  & 50  & 0.75 & MME1  \\
			15 & 1.5  & MME2  & 75  & 0.50 & MME1  \\
			20 & 1.0  & MME2  & 100 & 0.50 & MME1  \\
			\hline
		\end{tabular}
	} 
\end{table}
\section{{Asymptotic Behavior of the MLE}}

For the $SBFCD(\theta)$, the parameter space $\Theta = \mathbb{R} \setminus \{0\}$ is open and the \textit{pdf} (1.10) satisfies the conditions of Cramér-Rao inequality. Therefore, the standard asymptotic theory results apply to $\hat{\theta}_{ML}$, i.e., $\hat{\theta}_{\text{ML}} \xrightarrow{\text{P}} \theta$ and $ \sqrt{n I(\theta)} ( \hat {\theta}_{\text{ML}} - \theta ) \xrightarrow{\text{D}} N(0,1). $ In other words, for `large \textit{n}', $Bias(\hat{\theta}_{\text{ML}}) = E(\hat{\theta}_{\text{ML}}) - \theta) \approx 0 $ and
	\[
		M_n^*(\theta) = n [\text{MSE}(\hat{\theta}_{\text{ML}})] \approx  1/I(\theta), \tag{5.1}
	\]
where $I(\theta) = $ Fisher Information Per Observation (FIPO). But the expression of $I(\theta)$ is a bit complicated, and after a lengthy derivation it can be shown that 
	\[
		I(\theta) = I_1(\theta) -I_2(\theta), \tag{5.2} 
	\]	
where 
$ I_1(\theta) = \theta^{-2} +  \mathit{exp}(\theta)/(\mathit{exp}(\theta) - 1)^{2}; \text{ and } I_2(\theta) = 2E[J(U_1, U_2|\theta)], \text{ with } \\
J(U_1, U_2|\theta) = \left\{ J_1(U_1, U_2|\theta) / J_2(U_1, U_2|\theta) \right\}, \bm{U}=(U_1, U_2)^\prime \sim SBFCD(\theta),\text{ and}$
\begin{align*}
J_1(U_1, U_2|\theta) &= \mathit{exp}(-\theta(U_1 + U_2))\{-(U_1 - U_2)^2 + U_1^2 \mathit{exp}(-\theta U_2) + U_2^2 \mathit{exp}(-\theta U_1) + (U_1 + \\
&U_2 - 1)^2 \mathit{exp}(-\theta) \} + (U_2 - 1)^2 \mathit{exp}(-\theta(U_2 + 1)) + (U_1 - 1)^2 \mathit{exp}(-\theta(U_1 + 1)), \\
J_2(U_1, U_2|\theta) &= \left\{\mathit{exp}(-\theta U_1) + \mathit{exp}(-\theta U_2) - \mathit{exp}(-\theta) - \mathit{exp}\left(-\theta(U_1 + U_2)\right)\right\}^2.
\end{align*}

While the term $I_1(\theta)$ is easy to deal with, it is the term $I_2(\theta)$ which poses some challenge. It can be evaluated either by numerical integration as 
	\begin{align*}
	E[J(U_1, U_2|\theta)] &= \int_0^1 \int_0^1 J(u_1, u_2|\theta) \, c(u_1, u_2|\theta) \, du_1 \, du_2 = \text{a function of } \theta \text{ only}; \tag{5.3}
	\end{align*}
or by simulation as 
	\begin{align*}
	E[J(U_1, U_2|\theta)] &\approx (1/M) \sum_{m=1}^{M} J\left(U_1^{(m)}, U_2^{(m)} \mid \theta \right), \tag{5.4}
	\end{align*}
where $ (U_1^{(m)}, U_2^{(m)})^\prime = \bm{U}^{(m)}, \, 1 \leq m \leq M, \text{ are i.i.d. following } SBFCD(\theta).$ In the following we provide the plots of $I(\theta)$ and $1/I(\theta)$ in Figure 5.1.
	
\begin{figure}[H]
	\centering
	\begin{subfigure}[b]{0.48\linewidth}
		\centering
		\includegraphics[width=\linewidth]{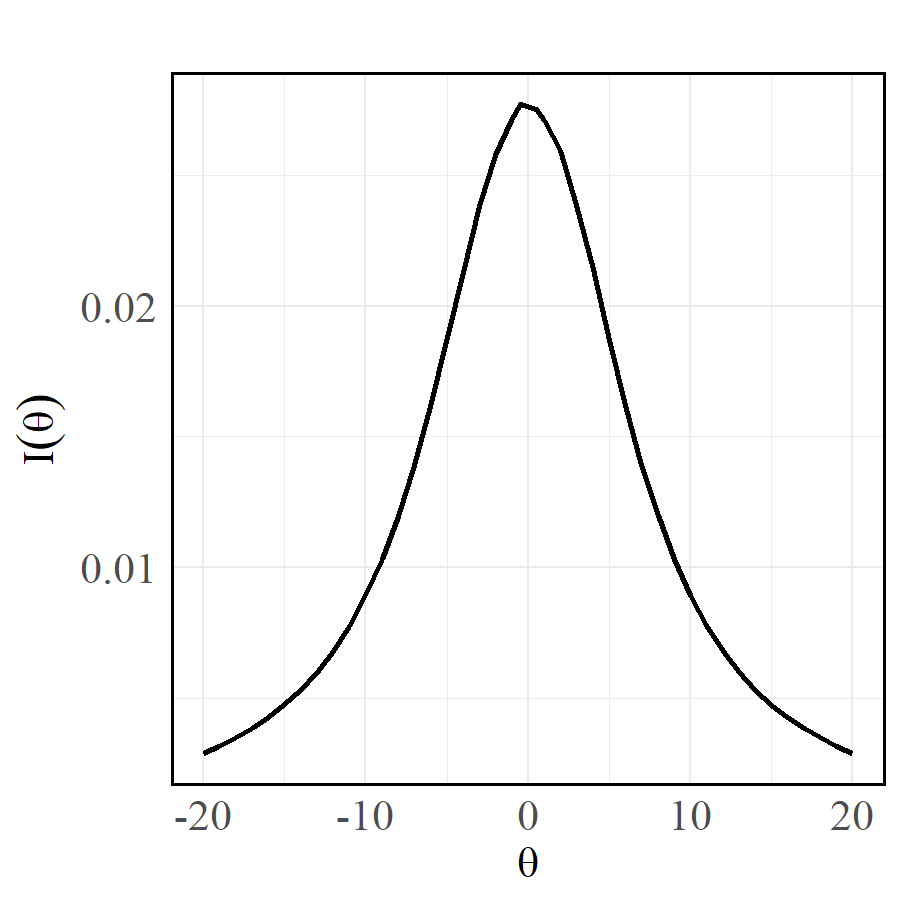}  
		\caption{}
		\label{fig:plot_l_theta}
	\end{subfigure}
	\hspace{0.02\linewidth}  
	\begin{subfigure}[b]{0.48\linewidth}
		\centering
		\includegraphics[width=\linewidth]{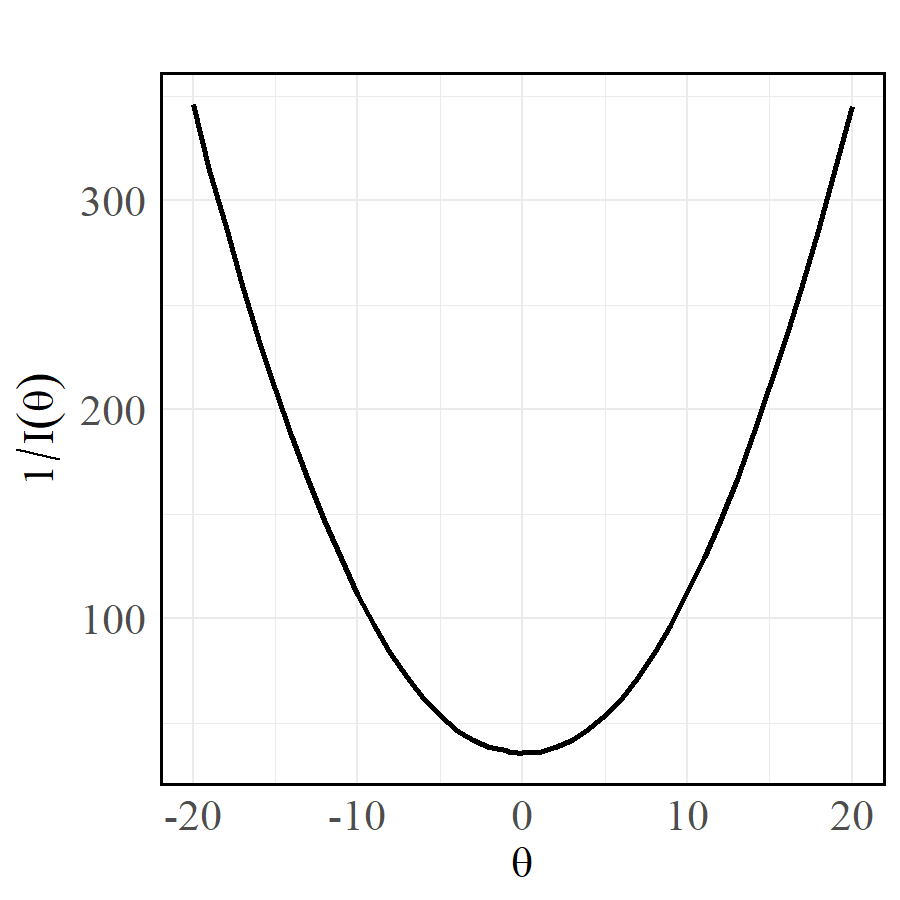}  
		\caption{}
		\label{fig:plot_H_theta}
	\end{subfigure}
	
	\caption{Figure 5.1. Plots of $I(\theta)$ (in (a)) and $1/I(\theta)$ (in (b)) as a function of $\theta$}
	\label{fig:plots_l_H_theta}
\end{figure}

	Since $\hat{\theta}_{ML}$ is first order efficient, i.e., $Bias(\hat{\theta}_{ML})$ is of order $O(n^{-1})$, $MSE(\hat{\theta}_{ML})$ and $Var(\hat{\theta}_{ML})$ are first order equivalent (i.e., they are same as far as the terms $O(n^{-1})$ are concerned). Hence, it is expected that
	\[
	M_n^*(\theta) = \left[ n \, \text{MSE}(\hat{\theta}_{ML}) \right] \to 1/ I(\theta), \quad \text{pointwise in} \, \theta, \, \text{as} \, n \to \infty. \tag{5.5}
	\] 
	
	The following Figure 5.2 shows the plots of $M_n^*(\theta) = \left[ n \, \text{MSE}(\hat{\theta}_{MLE}) \right]$ for $n = 25, 50, 75$ and $100$, along with $1/I(\theta)$. It is seen that $n = 75$ is good enough to attain the asymptotic bound $1/I(\theta)$.
	
	\begin{figure}[H]
		\centering
		\includegraphics[width=0.9\textwidth]{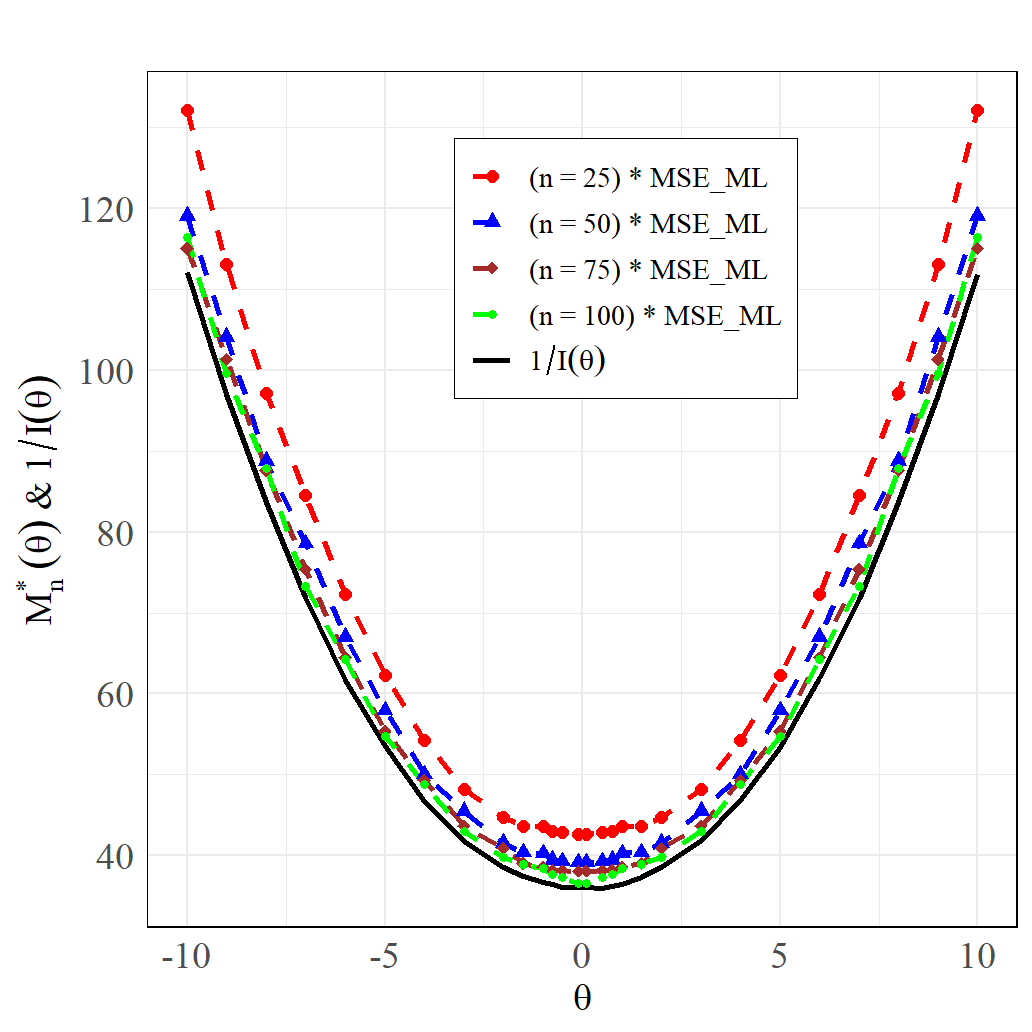} 
		\caption{Figure 5.2. Plots of $ M_n^*(\theta)$ for $n = 25, 50, 75$ and $100$ along with $1/I(\theta)$}
		\label{fig:a}
	\end{figure}
	
	The closeness of $M_n^*(\theta)$ to $1/I(\theta)$ has also been measured through the relative difference (RD) defined as
	\[
	RD = RD(\theta|n) = \left\{M_n^*(\theta) - 1/I(\theta)\right\} /M_n^*(\theta), \tag{5.6}
	\]
	and this has been shown in the following Table 5.1. By multiplying the values in Table 5.1 by $100$, we get the \% difference between the $\text{MSE}(\hat{\theta}_{ML})$ and asymptotic value $1/(nI(\theta))$.
		
	\renewcommand{\thetable}{5.1}
	
	\begin{table}[H]
		\centering
		\caption{Relative difference (RD) between $M_n^*(\theta)$ and $1/I(\theta)$.}
		\label{tab:rd_values}
		\begin{tabular}{@{\hskip 0.5cm}l@{\hskip 0.9cm}c@{\hskip 0.9cm}c@{\hskip 0.9cm}c@{\hskip 0.9cm}c@{}}
			\toprule
			$\theta$ & \text{RD ($n = 25$)} & \text{RD ($n = 50$)} & \text{RD ($n = 75$)} & \text{RD ($n = 100$)} \\ \midrule
		10 & 0.152 & 0.059 & 0.026 & 0.038 \\
		9  & 0.140 & 0.067 & 0.041 & 0.024 \\
		8  & 0.138 & 0.057 & 0.045 & 0.044 \\
		7  & 0.152 & 0.088 & 0.050 & 0.023 \\
		6  & 0.144 & 0.077 & 0.040 & 0.038 \\
		5  & 0.147 & 0.083 & 0.040 & 0.028 \\
		4  & 0.138 & 0.064 & 0.051 & 0.042 \\
		3  & 0.131 & 0.081 & 0.042 & 0.026 \\
		2  & 0.131 & 0.065 & 0.051 & 0.024 \\
		1.5 & 0.137 & 0.068 & 0.035 & 0.032 \\
		1   & 0.149 & 0.087 & 0.047 & 0.041 \\
		0.75 & 0.150 & 0.074 & 0.044 & 0.031 \\
		0.5  & 0.151 & 0.073 & 0.046 & 0.026 \\
		0.1  & 0.150 & 0.075 & 0.049 & 0.009 \\ \bottomrule
		\end{tabular}
	
	\end{table}
	
\noindent \textbf{Remark 5.1.} The above Table 5.1 shows how close the MLE is in terms of attaining its asymptotic MSE for various values of $n$. For $n=25$, it comes within $15\%$ of its target value; for $n=50, 75$ and $100$, it comes within $9\%, 5\%$ and $4\%$ of $1/(nI(\theta))$, respectively. Using the $5\%$ threshold we belive that $n=75$ is good enough for the MLE to attain its limiting variance (and the MSE). Also, values of $\theta$ outside $\{|\theta| \leq 10 \}$ makes no further difference (though not reported here for brevity).
				
\section{Conclusion}
This work provides a comprehensive study of the point estimation of the association parameter of a bivariate Frank copula which, to the best of our knowledge, was not reported before. Three point estimators have been studied extensively in terms of bias and MSE, and it has been found that even though the MLE has the best overall performance, the other two estimators (MMEs) can have an advantage in terms of MSE near the origin. It has also been demonstrated that the MLE attains its asymptotic variance (and MSE) fairly well for a sample size as low as $75$. Going forward, our next objective will be to study the hypothesis testing problem, especially for testing $\theta = 0$ (i.e., independence) where the above results may be useful. 

\newpage
\section*{Acknowledgements}

	The authors would like to express their sincere gratitude to Ton Duc Thang University (TDTU), Ho Chi Minh City, and Ho Chi Minh City University of Technology (HCMUT) for providing favorable conditions for this collaborative research. The logistical help and encouragement extended by both the institutions have played a crucial role in the completion of this research.
	Also, this work was initiated while the authors were visiting the Vietnam Institute for Advanced Study in Mathematics (VIASM) in the summer of 2024. The authors would like to thank the VIASM administration for their generous support and hospitality which accelerated the progress of this work and contributed in achieving the research results.
	
\bibliographystyle{plain}

\addcontentsline{toc}{section}{References}
	
	\begin{thebibliography}{99}
		
		\bibitem{Chatterjee2022} Chatterjee, R. (2022). \textit{Inferences for the Bivariate Probability Distribution Using Farlie-Gumbel-Morgenstern Copula}. PhD thesis. Available at: \url{https://www.proquest.com/dissertations-theses/inferences-bivariate-probability-distribution/docview/2882152890/se-2}. 
		
		\bibitem{Dette2014} Dette, H., Hecke, R. V., and Volgushev, S. (2014). Some Comments on Copula-Based Regression. \textit{Journal of The American Statistical Association}, Vol. 109, No. 507, 1319-1324.
		
		\bibitem{Fairuz2024} Fairuz, N. S. N. M., Norrulashikin, S. M., Pahrany, A. D., and Kamisan, N. A. B. (2024). Parameter Estimation Methods for Bivariate Copula in Financial Application. \textit{Semarak International Journal of Modern Accounting and Finance}, Vol. 2, No. 1, 10-21.
		
		\bibitem{Genest1987} Genest, C. (1987). Frank’s Family of Bivariate Distributions. \textit{Biometrika}, 74, 549-555. \url{https://doi.org/10.1093/biomet/74.3.549}.
		
		\bibitem{Jouanin2004} Jouanin, J-F., Riboulet, G., and Roncalli, T. (2004). Financial Applications of Copula Functions in Risk Measures For The 21st Century. In: Par Giorgio Szego (Ed.), \textit{John Wiley \& Sons}. Available at SSRN: \url{https://ssrn.com/abstract=1032588}.
		
		\bibitem{Nelsen2007} Nelsen, R. B. (2007). \textit{An Introduction to Copulas}. Springer Science \& Business Media.
		
		\bibitem{Sklar1959} Sklar, M. (1959). Fonctions de repartition an dimensions et leurs marges. \textit{Publ. Inst. Statist. Univ. Paris}, 8, 229–231.
		
	\end{thebibliography}

	\newpage
	
	\appendix
	\section*{Appendix} 

	\textbf{Appendix A.1:} \textbf{Existing R Code for Generating Sample Pairs $(U_1, U_2)$} 


\begin{lstlisting}
	# Number of samples
	n <- 100
	
	# Initialize a matrix to store pairs (U1, U2)
	sample_U1U2 <- matrix(nrow = n, ncol = 2)
	
	# Set seed for reproducibility
	# set.seed(123)
	
	# Loop to generate samples
	for (i in 1:n) {
		u1 <- runif(1, 0, 1)
		v <- runif(1, 0, 1)
		theta <- 10  # Adjust this value as needed
		A <- (exp(-theta * u1) - exp(-theta))
		B <- (1 - exp(-theta * u1))
		D <- (1 - exp(-theta))
		T1 <- D * (A / B + 1)
		T2 <- (v * (exp(theta * u1) - 1) * (A + B) + D)
		u2 <- (-1 / theta) * log((T1 / T2) - (A / B))
		
		# Store the pair (U1, U2) in the matrix
		sample_U1U2[i, ] <- c(u1, u2)
	}
	
	# Print the sample
	print("Generated sample of (U1, U2) pairs:")
	print(sample_U1U2)
\end{lstlisting} 
\vspace{0.5cm}  
\textbf{Appendix A.2:} \textbf{Generating a value $ \mathbf{u} =(u_1,u_2)$ of $\mathbf{U}$ from the joint \textit{pdf} (1.10) }\\ [0.5cm]
	Generate $U_1 =u_1$ following $Uniform(0,1)$. The marginal \textit{pdf} value of $U_1$ at $u_1$ is 1.\\
	For the above generated $U_1 =u_1$, the conditional \textit{pdf} of $(U_2 = v_* \mid U_1 =u_1) $=  \{joint  \textit{pdf} of $(U_1,U_2)$ at $(u,v_*)$ using (1.10)\} = $c_{\theta}((U_2 = v_* \mid U_1 =u_1))$ (say)\\
	Hence the conditional \textit{cdf} of $(U_2 \mid U_1 =u_1)$ at $u_2$ is \\
	
$= \int_0^{u_2} c_\theta(U_2 = v_* \mid U_1 = u_1) \, dv_*$

\[
    = \theta (1 - \mathit{exp}(-\theta)) \mathit{exp}(-\theta u_1) 
\int_0^{u_2} \mathit{exp}(-\theta v_*) \big[ \mathit{exp}(-\theta u_1) + \mathit{exp}(-\theta v_*) -  \mathit{exp}(-\theta)
\]
		\[
		 - \mathit{exp}(-\theta u_1 - \theta v_*) \big]^{-2} \, dv_*
		\]
		
$ = C_\theta(U_2 = u_2 \mid U_1 = u_1) \text{(say)}\in (0, 1) $

\vspace*{0.5cm}

So, now generate \( v \sim \textit{Uniform}(0, 1) \) which can be equated with the above \( C_\theta(U_2 = u_2 \mid U_1 = u_1) \), and now by solving for \( u_2 \) we get the conditional value of \( U_2 \) for given \( U_1 = u_1 \). This is what has been shown in Subsection 1.4.\\ [0.5cm]
\textbf{Appendix A.3:} \textbf{A Random Sample of size $\bm{25}$} 
	\renewcommand{\thetable}{A.1} 
	
	\begin{table}[h!]
		\centering
		\caption{A random sample of size $n = 25$ observations
			from SBFCD($\theta$) with $\theta = 1$}
		
		\setlength{\tabcolsep}{9pt} 
			
		\resizebox{0.95\textwidth}{!}{ 

			\begin{tabular}{|c| c c || c| c c || c| c c|}
				\hline
				\textbf{ID} & \textbf{U1} & \textbf{U2} & \textbf{ID} & \textbf{U1} & \textbf{U2} & \textbf{ID} & \textbf{U1} & \textbf{U2} \\
				\hline
				1  & 0.2876 & 0.7468 & 9  & 0.2461 & 0.0342 & 17 & 0.6907 & 0.8209 \\
				2  & 0.4090 & 0.8699 & 10 & 0.3279 & 0.9445 & 18 & 0.0246 & 0.3652 \\
				3  & 0.9405 & 0.0713 & 11 & 0.8895 & 0.7651 & 19 & 0.7585 & 0.2672 \\
				4  & 0.5281 & 0.8920 & 12 & 0.6405 & 0.9948 & 20 & 0.3182 & 0.2048 \\
				5  & 0.5514 & 0.4700 & 13 & 0.6557 & 0.7358 & 21 & 0.1428 & 0.3343 \\
				6  & 0.9568 & 0.5657 & 14 & 0.5441 & 0.6027 & 22 & 0.4137 & 0.3518 \\
				7  & 0.6776 & 0.6132 & 15 & 0.2892 & 0.1259 & 23 & 0.1524 & 0.1053 \\
				8  & 0.1029 & 0.8543 & 16 & 0.9630 & 0.9340 & 24 & 0.2330 & 0.4025 \\
				&        &        &    &        &        & 25 & 0.2660 & 0.8230 \\
				\hline
			\end{tabular}
		}
	\end{table}

\end{document}